\documentclass[aps,prd,preprint,superscriptaddress,nofootinbib,showpacs]{revtex4}
\usepackage{graphicx}
\usepackage{color}
\usepackage[normalem]{ulem}
\usepackage{xcolor}

\definecolor{My_red}        {cmyk}{0.00,1.00,1.00,0.20}

\newcommand{\bmat}{\left(\begin{array}}
\newcommand{\emat}{\end{array}\right)}
\newcommand{\beq}{\begin{equation}}
\newcommand{\eeq}{\end{equation}}
\newcommand{\wt}{\widetilde}

\def\ra{\rightarrow}

\def\ld{\lambda}
\def\f{\frac}

\def\bwt{\begin{widetext}}
\def\ewt{\end{widetext}}
\def\be{\begin{equation}}
\def\ee{\end{equation}}
\def\bary{\begin{array}}
\def\eary{\end{array}}
\def\bit{\begin{itemize}}
\def\eit{\end{itemize}}
\def\GeV{\rm GeV}

\def\ra{\rightarrow}

\def\ol{\overline}

\def\ld{\lambda}

\def\su5u1{SU(5) \times U(1)}
\def\fsu5u1{SU(5) \times U(1)'}
\def\so10{SO(10)}
\def\sq20{SO(10) \times SO(10)}

\def\ra{\rightarrow}

\def\ld{\lambda}
\def\f{\frac}

\def\L{\left(}
\def\R{\right)}
\def\bwt{\begin{widetext}}
\def\ewt{\end{widetext}}
\def\be{\begin{equation}}
\def\ee{\end{equation}}
\def\bary{\begin{array}}
\def\eary{\end{array}}
\def\bit{\begin{itemize}}
\def\eit{\end{itemize}}
\def\GeV{\rm GeV}

\def\ra{\rightarrow}

\def\ol{\overline}

\def\ld{\lambda}

\def\su5u1{SU(5) \times U(1)}
\def\fsu5u1{SU(5) \times U(1)'}
\def\so10{SO(10)}
\def\sq20{SO(10) \times SO(10)}

\usepackage[centertags]{amsmath}
\usepackage{amssymb}

\begin{document}

\title{Naturally Large Radiative Lepton Flavor Violating Higgs Decay Mediated by Lepton-flavored Dark Matter}

\author{Seungwon Baek}
\email{swbaek@kias.re.kr}
\affiliation{School of Physics, KIAS, 85 Hoegiro Dongdaemun-gu, Seoul 02455, Korea}

\author{Zhaofeng Kang}
\email{zhaofengkang@gmail.com}
\affiliation{School of Physics, KIAS, 85 Hoegiro Dongdaemun-gu, Seoul 02455, Korea}


\date{\today}

\begin{abstract}

In the standard model (SM), lepton flavor violating (LFV) Higgs decay is absent at renormalizable level and thus it is a good probe to new physics. In this article we study a type of new physics that could lead to large LFV Higgs decay, i.e., a lepton-flavored dark matter (DM) model which is specified by a Majorana DM
  and scalar lepton mediators. Different from other similar models with similar setup, we introduce both left-handed and right-handed scalar leptons. They allow large LFV Higgs decay and thus may explain the tentative Br$(h\ra\tau\mu)\sim1\%$ experimental results from the LHC. In particular, we find that the stringent bound from $\tau\ra\mu\gamma$ can be naturally evaded. One reason, among others, is a large chirality violation in the mediator sector. Aspects of relic density and especially radiative direct detection of the leptonic DM are also investigated, stressing the difference from previous lepton-flavored DM models.

\end{abstract}

\maketitle

\section{Introduction and Motivation}

After the discovery of standard model (SM) Higgs-like boson, the next step is to measure its couplings precisely to see possible deviation from the SM and thus to search for new physics. The Yukawa couplings between Higgs boson and charged leptons that cause lepton flavor violation (LFV) are of particular interest, because in the SM they are absent at tree level and highly suppressed at loop levels, and thus are sensitive to new physics. As a matter of fact, in most of models that address neutrino masses and oscillations, LFV is well expected and has already been observed in neutrino oscillations described by the PMNS matrix. Moreover, although charged lepton flavor violation (CLFV) has not been observed yet, in general those models should leave measurable signals in processes like $\mu\ra e\gamma$, $\mu\ra 3e$, etc. A lot of efforts have been devoted to searching  for CLFV and the null results impose very strong bounds on the magnitude of LFV~\cite{Bernstein:2013hba}. 

Searching for LFV Higgs decays~\cite{HLFV:early} receives special attention in the LHC era~\cite{Harnik:2012pb}. The CMS collaboration reported the upper limit Br$(h\ra \tau\mu)<1.57\%$ at 95$\%$ C.L., using the 19.7 fb$^{-1}$ of $\sqrt{s} = 8$ TeV data~\cite{Khachatryan:2015kon}. Interestingly, the best fit (assuming both the production cross section and total width of Higgs being SM-like) hints a 2.4 $\sigma$ excess with Br$(h\ra \tau\mu) = (0.84^{+0.39}_{-0.37})\%$. More recently, the ATLAS collaboration obtained an the upper limit Br$(h\ra \tau\mu)<1.85\%$ from hadronic $\tau$ decay at 95$\%$ C.L., using the 20.3 fb$^{-1}$ of $\sqrt{s} = 8$ TeV data~\cite{LFV:ATLAS}. Although they have not seen significant deviation from the SM, their best fit value Br$(h\ra \tau\mu) = (0.77\pm 0.62)\%$ is consistent with the CMS { result}. At the 300 fb$^{-1}$ of 13 TeV LHC, the sensitivity can reach { down to} 7.7$\times10^{-4}$ and thus the CMS excess will be confirmed or excluded~\cite{Kopp:2014rva}. 

In the models with canonical seesaw mechanism LFV Higgs decay is too small to be observed~\cite{Pilaftsis,Arganda:2004bz}. { This is because of} the decoupling of right-handed neutrinos (RHNs) either through the smallness of Yukawa couplings or heaviness of RHNs. In the inverse seesaw mechanism, where sizable Yukawa couplings are allowed for light RHNs, appreciable LFV Higgs decay can be accommodated~\cite{Arganda:2014dta,Arganda:2015naa}. 

Alternatively, the tiny neutrino masses can be generated by radiative corrections~\cite{Loop:nu1,Loop:nu2}. However, to our knowledge, none of { those radiative seesaw models} could generate large LFV Higgs decay. Actually, facing the stringent constraint from CLFV, it is quite nontrivial to get LFV Higgs decay { large enough to detect at the LHC}. 

At tree level, two (or even more)-Higgs doublet model (2HDM) with proper flavor changing neutral current allows LFV Higgs decay which is large enough to explain the CMS excess~\cite{Vicente:2014qea,Heeck:2014qea,Crivellin:2015mga,Dorsner:2015mja,Omura:2015nja,Mao:2015hwa,Botella:2015hoa,Campos:2014zaa,deLima:2015pqa,Liu:2015oaa,Crivellin:2015hha}. Higher { dimensional} operators in the effective theory framework { were also considered}~\cite{deLima:2015pqa,Dorsner:2015mja,He:2015rqa,Altmannshofer:2015esa}. But at loop level a large cancellation probably is needed to evade the CLFV constraint~\cite{Dorsner:2015mja,Cheung:2015yga,Baek:2015mea}. { Other scenarios can be found in Ref.~\cite{Chiang:2015vpt,Huang:2015vpt}.}

In this article we establish a connection between LFV Higgs decay and { a} type of dark matter (DM), {\it i.e.}, lepton-flavored DM~\cite{Bi:2009md,Agrawal,Lee:2014rba,Hamze:2014wca,Kile:2014jea,Bai:2014osa,Chang:2014tea,Geng:2014zqa,Agrawal:2015tfa}. { In this scenario, DM interacts merely with the SM lepton sector, whereupon DM-quark interactions arise at loop level.}  
An obvious merit of { that} kind of DM is { that we can  easily understand the} null results from DM direct detection experiments such as LUX~\cite{Akerib:2013tjd}. 

That paradigm can be achieved in two ways. One way is introducing a leptophilic vector boson or Higgs boson propagating in the $s-$channel { for the DM pair annihilation diagrams}. This kind of model gives rise to poor flavor { phenomenology}. 

The other way is introducing mediators in the $t-$channel to form lepton flavored DM~\footnote{Note that in this way the leptonic nature of DM is naturally specified by the quantum numbers of mediators with respect to SM. No extra local or global leptonic symmetry is required. }. Then, LFV { can} happen in the dark sector and is mediated to the SM sector via loop processes. 
Furthermore, mediators could consist of both left-handed and right-handed scalar leptons (the previous studies { were} based on only one type of them), just { as in the case of the supersymmetric SMs}. Remarkably, we find that this kind of lepton-flavored DM is able to accommodate LFV Higgs decay while other models with only one type chirality fail to. As an example, { we will show} that in our model a sizable Br$(h\ra \tau\mu)$ at { the} level { of} $1\%$ can be naturally achieved without incurring too large Br($\tau\ra \mu\gamma$). It is attributed { partly} to the large chirality flipping in the scalar sector and { also to} the cancellation between { different} contributions to CLFV. In addition, we study the { mechanism} for DM, a Majorana fermion, to acquire correct relic density. For the weak scale DM, even $s-$wave annihilation may work without large Yukawa couplings. Related to radiative LFV Higgs decay, radiative correction could also lead to Higgs-mediated DM-nucleon scattering which may be detected in the near future.

 This paper is organized as follows:  In Section~\ref{sec:model}  the model is introduced. 
In Section~\ref{sec:LFV} we consider Higgs LFV decay confronting charged lepton LFV decay, along with others. 
In Section~\ref{sec:dm} we study { relic density and direct detection of our leptophilic} dark matter and their relations with LFV Higgs decay. { We conclude in Section~\ref{sec:concl}.}

\section{Lepton-flavored Majorana dark matter }
\label{sec:model}

In this section we will first present the model in its simplest version, and then calculate the mass spectra that will be used later.

\subsection{The model with dual mediators}

 { From} model building perspective, { a} natural way to realize a lepton-flavored DM is to introduce
a Majorana DM candidate connected to some lepton flavors by means of scalar leptons. 
If DM is a scalar field, whether it is real or complex, it is hard to get rid of the conventional Higgs portal in a natural manner, 
not to mention other demerits. So we focus on the case where DM is a singlet Majorana { fermion} 
{ $N$}, protected by a $Z_2$ dark matter parity. 
At the renormalizable level,   { $N$} can not couple to SM fields. Its interactions with SM fields necessitate additional mediators, 
and we can specify these interactions by introducing mediators with proper quantum numbers. 
In order to make up a lepton-flavored DM, one can designate a scalar partner for { each} SM { left-handed} lepton { doublet} $l_L$ and { right-handed lepton signlet} $e_R$. 
They are labelled as $\phi_\ell$ and $\phi_e$, respectively. For simplicity, only a single family of scalar lepton (slepton for short, borrowing the name 
from supersymmetry) will be considered. In this paper we do not have the ambition to address the flavor structure of the 
dark sector by imposing flavor symmetry. We just treat all the couplings as free parameters.

With the degrees of freedom at hand, restricted by the $Z_2$ dark matter parity under which only the new particles are odd, the most general Lagrangian (aside from the kinetic energy terms) takes a form of
\begin{eqnarray}\label{Lag}
-{\cal L}=&-{\cal L}_{\rm SM}+{ m^2_{\phi_l} |\phi_\ell|^2 + m^2_{\phi_e} |\phi_e|^2 +{1 \over 2} M \ol{N} N }+\L -y_{La}\bar l_a  P_R
              { N} \wt \phi_\ell+y_{Ra}\bar e_a P_L  { N} \phi_e+h.c.\R\cr
&+\L -\mu\,H^\dagger\wt \phi_\ell  \phi_e^*+h.c.\R+\ld_{-1}|\phi_e|^2|\phi_\ell|^2+\ld_{0} |H|^2|\phi_e|^2+V_{\rm 2HDM},
\end{eqnarray}  
where $\wt\phi_\ell\equiv i\sigma_2\phi_\ell^*$. In our convention $\phi_\ell$ is assigned with { the same} hypercharge $Y=+1/2$ with the SM Higgs doublet 
$H$ so that $\phi_\ell$ can be regarded as the 2nd Higgs doublet in 2HDM. 
Couplings $\ld_{-1}$ and $\ld_0$ are not important in our ensuing discussions and are  { set} to be zero. 
The part involving the two Higgs doublets, as usual, is given by 
\begin{align}\label{}
V_{\rm 2HDM}=\frac{\ld_1}{2}|\phi_\ell|^4+\frac{\ld_2}{2}|H|^4+\lambda_3|\phi_\ell|^2|H|^2+\lambda_4\left(\phi_\ell^\dagger{H}\right)\left({H}^\dagger\phi_\ell\right)
+\L \frac{\ld_5}{2}\left(\phi_\ell^\dagger{H}\right)^2+h.c.\R.
\end{align}  
In this potential most parameters are irrelevant to { our} phenomenological studies, except for $\ld_5$ that is crucial in neutrino mass generation.

A comment deserves special attention. We start from lepton-flavored DM, but as a bonus nonzero neutrino masses { are generated as} a generic consequence
of this type of DM model. It is obvious that all of the crucial ingredients of the Ma's model~\cite{Loop:nu2} are incorporated in our framework, 
and thus radiative corrections lead to neutrino masses:
\begin{align}\label{Loop:nu2}
m_\nu \sim \ld_5 \f{y_{La}^2}{16\pi^2} \L\f{v}{m_{\phi_\ell}}\R^2 M,
\end{align}
with $v=246$ GeV. In the parameter space relevant to LFV Higgs decay, $M$ is around the weak scale while $m_{\phi_\ell}\sim{\cal O}(\rm TeV)$ 
and moreover $y_{La}\sim {\cal O}(1)$. Then the resulting neutrino mass scale is much above the eV scale except for extremely suppressed 
$\ld_5\ll1$. In this paper we will not pay further attention on this aspect and always assume a sufficiently small $\ld_5$ to suppress radiative neutrino mass.

\subsection{The mass spectrum of the mediators}

\begin{figure}[htb]
\includegraphics[width=3.0in]{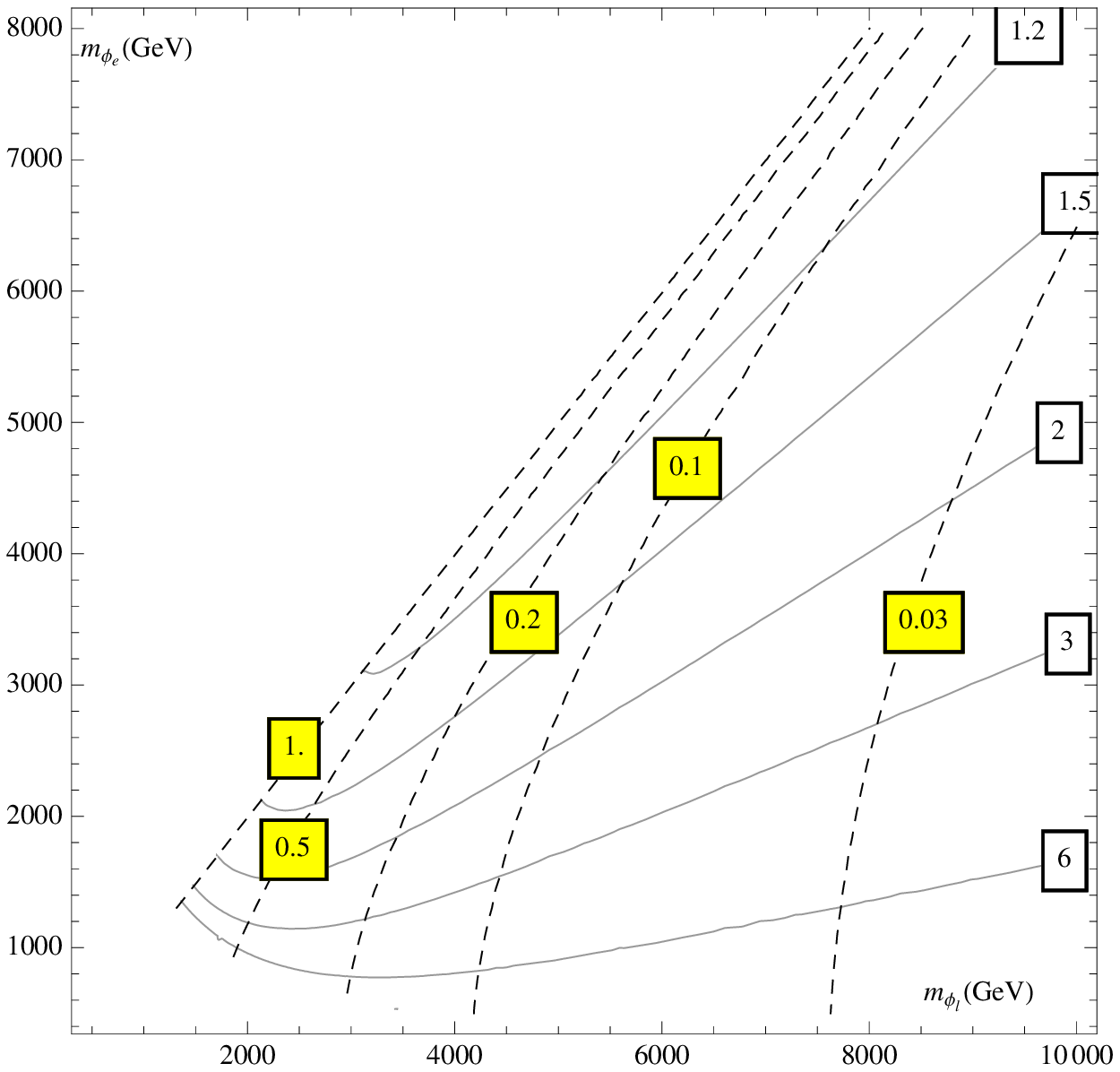}
\includegraphics[width=3.0in]{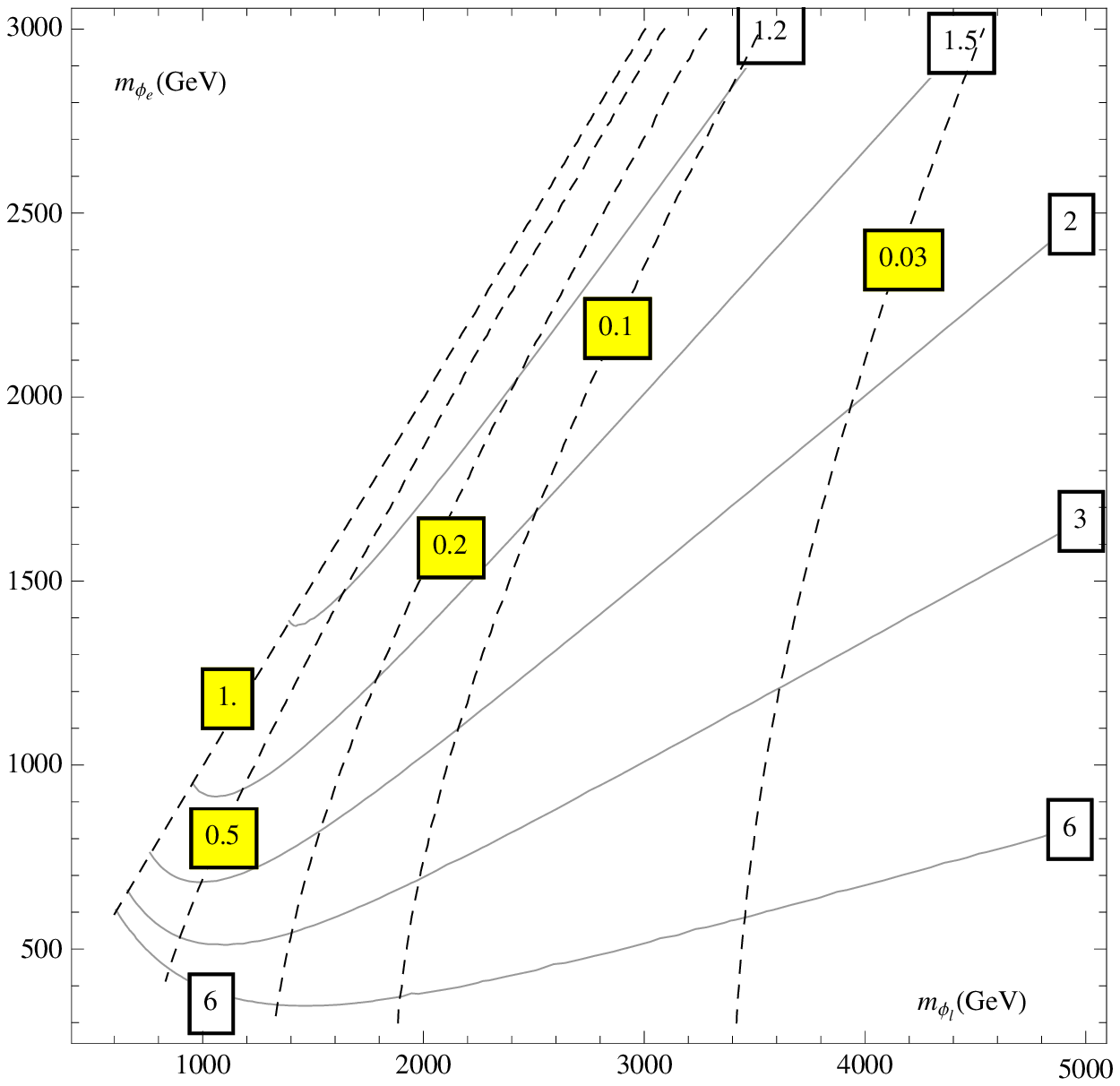}
\caption{Contour plots of $m_{\wt e_2}/m_{\wt e_1}$ (solid lines) and $\tan\theta$ (solid lines) on the $m_{\phi_\ell}-m_{\phi_e}$ plane: left $\mu=10$ TeV; right $\mu=2$ TeV. }\label{mixing}
\end{figure}

In the right vacuum, only $H$ is supposed to develop vacuum expectation value (VEV), breaking the electroweak symmetries but not $Z_2$. 
Then the charged component of $\phi_\ell$, which is written in component as $\phi_\ell=(\phi_\ell^+,(\phi_{R}+i\phi_I)/\sqrt{2})^T$, 
would mix with $\phi_e$ through the $\mu-$term, i.e., $\mu v \phi_e\phi_\ell^+/\sqrt{2}+c.c.$. Then mass eigenstates are related to the flavor eigenstates
via
{
\be\label{}
\wt e_1=\cos\theta (\phi_\ell^+)^*-\sin\theta\, \phi_e,\quad \wt e_2=\sin\theta (\phi_\ell^+)^*+\cos\theta\, \phi_e,
\ee
}
The two charged sleptons respectively have the following (mass)$^2$
\be\label{}
m_{\wt e_{1,2}}^2=\f{1}{2}\left[ \L m_{\phi_\ell}^2+m_{\phi_e}^2\R \mp\sqrt{\L m_{\phi_\ell}^2-m_{\phi_e}^2\R^2+2\mu^2v^2}\right],
\ee
respectively.
The $\ld_3-$term contributions to masses have been absorbed into the bare mass term of $\phi_\ell$,  $m_{\phi_\ell}^2$
which is common to all components. 
And  { similar} operation is { done} for $\phi_e$. The mixing angle,
within  { $(-\pi/2,\pi/2)$}, is given by
\be\label{theta}
\tan\theta=\f{1}{\sqrt{2}\mu v}\left[ \L m_{\phi_\ell}^2-m_{\phi_e}^2\R+ \sqrt{\L m_{\phi_\ell}^2-m_{\phi_e}^2\R^2+2\mu^2v^2}\right].
\ee
For completeness, we also give masses for the two neutral components. Their mass degeneracy is lifted by terms in the $V_{\rm 2HDM}$, 
 \be\label{mass:sp}
m_{\phi_R}^2\approx m_{\phi_\ell}^2+\L\ld_4+\ld_5\R v^2/2,\quad m_{\phi_{ I}}^2\approx m_{\phi_\ell}^2+\L\ld_4-\ld_5\R v^2/2.
\ee 
For future convenience, in Fig.~\ref{mixing} we show the mass ratio $m_{\wt e_2}/m_{\wt e_1}$ and $\tan\theta$ for the cases with a very large and normal $\mu$, respectively.

It is useful to expand the Lagrangian Eq.~(\ref{Lag}) in components. For a more general setup, we introduce a Lagrangian that contains a couple of scalar fields $\phi_\ell^+$ with unit charge and as well several Majorana fermion $N_\alpha$ instead of only one (for example in the Ma's model there are three RHNs). Their interactions in the mass basis are given by 
\begin{align}\label{model:com}
-{\cal L}=&m_{\wt e_i}^2|\wt e_i|^2+ { \f{M_\alpha}{2}\ol
            N_\alpha N_\alpha} +\f{1}{2}m_h^2h^2\cr
&
+A_{ij} h\wt e_i^*\wt e_j+\left[\wt e_i \bar e_{a}\L \ld_{ia\alpha}^L P_L+\ld_{ia\alpha}^R P_R\R N_\alpha+h.c.\right],
\end{align}  
with $a=1,2,3$ the { generation} index. It is assumed that $N_\alpha$'s are Majorana fermions, but practically this assumption is not necessary 
for generating LFV Higgs decay (but necessary for generating neutrino masses). Expressed in terms of the original parameters, 
the couplings can be written as
\begin{align}\label{}
&{ A_{11}=-A_{22}=-\f{\mu}{\sqrt{2}}\sin2\theta,\quad A_{12}=A_{21}=\f{\mu}{\sqrt{2}}\cos2\theta,}\\
&{ \ld^L_{1a\alpha}=-\sin\theta y_{Ra\alpha},\quad \ld^L_{2a\alpha}=\cos\theta y_{Ra\alpha};\quad
  \ld^R_{1a\alpha}=\cos\theta y_{La\alpha},\quad \ld^R_{2a\alpha}=\sin\theta y_{La\alpha}}, 
\end{align}  
The two neutral sleptons do not play important roles in the { following} discussions because they do not couple to the Higgs boson with a large massive coupling.

\section{$h\ra \tau\mu$ confronting $\tau\ra \mu\gamma$ and  $h\ra \gamma\gamma$}
\label{sec:LFV}

In this section we will investigate how to get large LFV Higgs decay without { conflict} with the strong constraints such as CLFV or $h\ra\gamma\gamma$. We will concentrate on $h\ra \tau\mu$ as an example, but the discussions can be applied to other similar processes.

\subsection{Radiative LFV Higgs decay: $h\ra\bar \ell_a\ell_b$}

\begin{figure}[htb]
\includegraphics[width=1.5in]{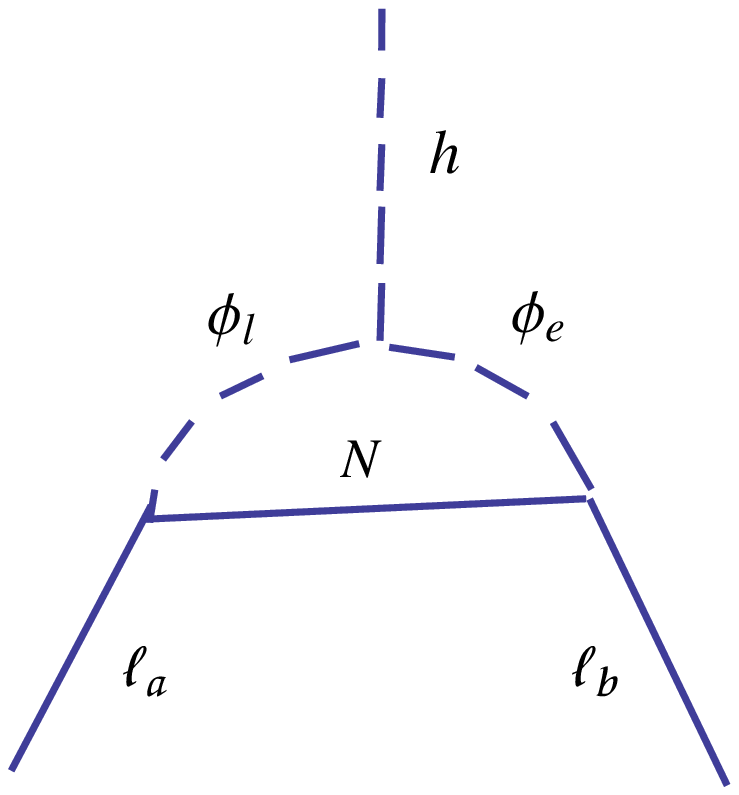}~
\includegraphics[width=1.5in]{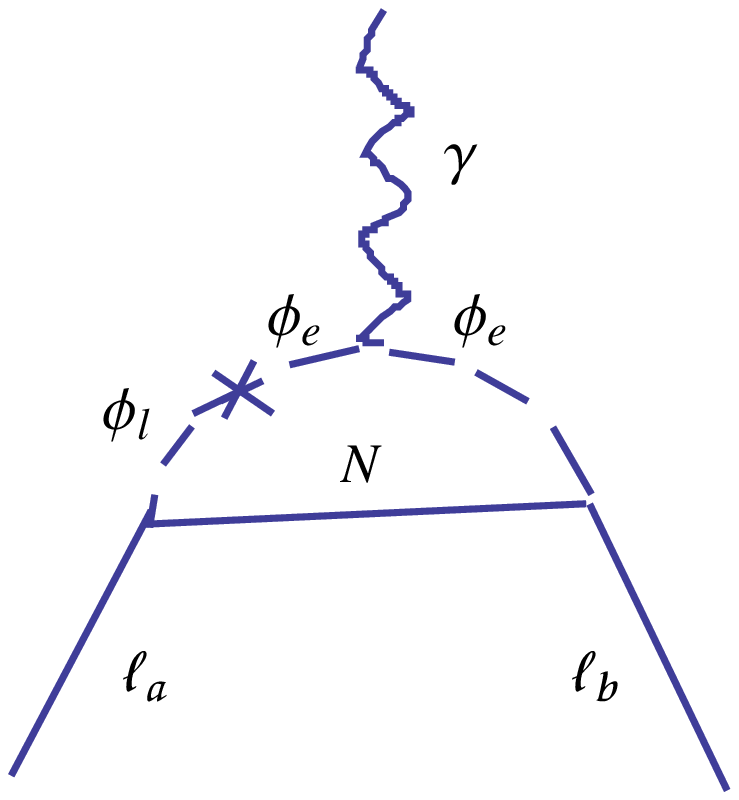}~
\includegraphics[width=1.4in]{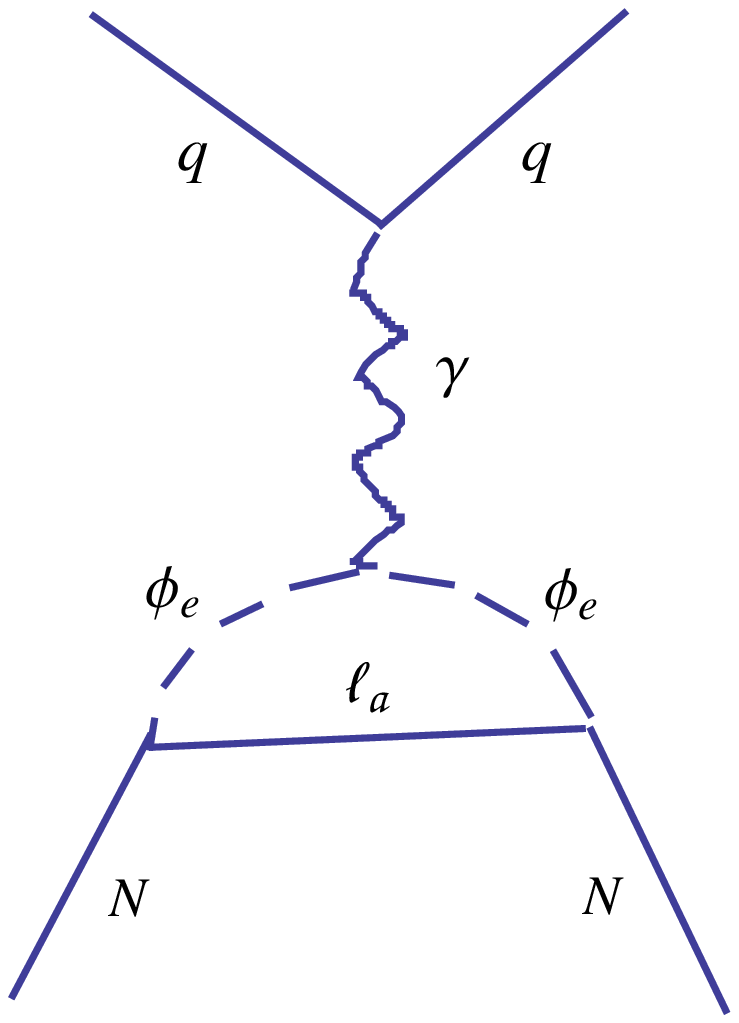}~
\includegraphics[width=1.4in]{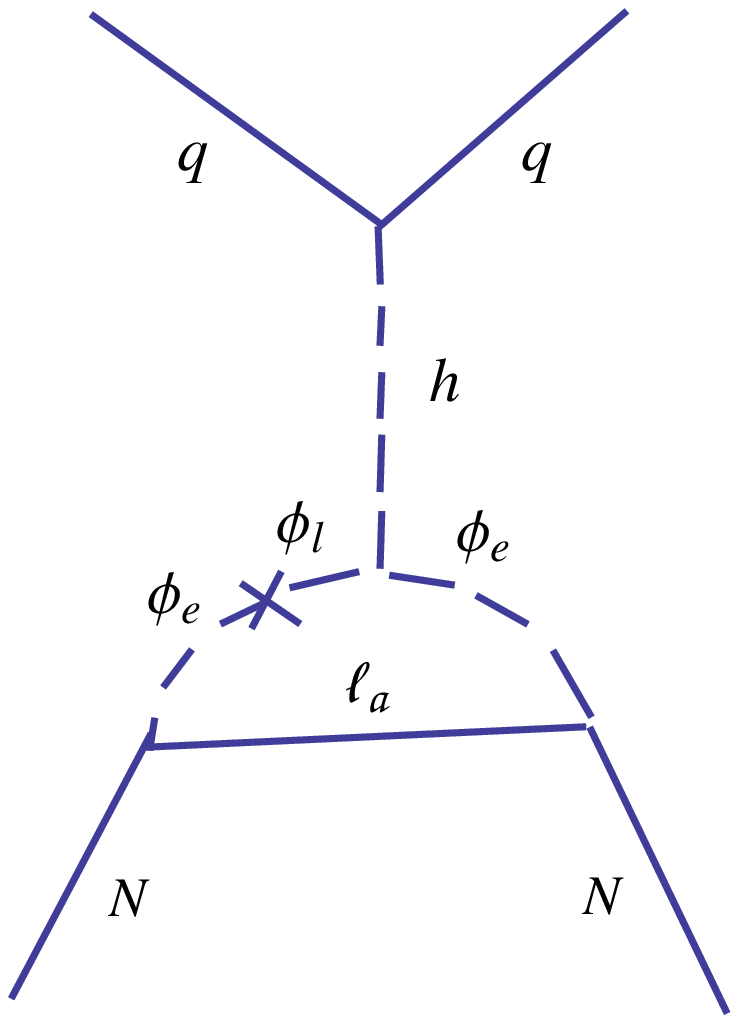}
\caption{Schematic Feymann diagrams for Higgs (first panel) and charged lepton (second panel) LFV decays; Photon (third panel) and Higgs (fourth panel) mediated DM-quark scattering. Loopy particles are in the interacting basis to manifest the dependence on mixing.}\label{radiative}
\end{figure}

The charged sleptons $\wt e_i$ mediate radiative Higgs LFV decay $h\ra\bar \ell_a\ell_b$, with the Feymann diagram { shown} in the first panel of Fig.~\ref{radiative}. The corresponding amplitude is generically written as
\begin{align}\label{eq:h2mt}
i{\cal M}=+ i \bar u_b(-p_2+p_1) \L F_L P_L+F_R P_R\R v_a(p_2),  
\end{align}
with the form factor 
\begin{align}\label{FL}
F_L=&\f{1}{16\pi^2}M_\alpha C_0(-p_2,p_1-p_2,M_\alpha,m_{\wt e_i},m_{\wt e_j}) A_{ji}\L\ld_{ia\alpha}^{R}\R^*\ld_{jb\alpha}^L\cr
\simeq&\f{1}{16\pi^2} \f{\mu}{\sqrt{2}M_\alpha} y_{Rb\alpha}y^*_{La\alpha}\left[ \f{1}{2} \sin^22\theta \L G(x_1)+G(x_2)\R
+\cos^22\theta\, G(x_1,x_2) \right],
\end{align}
{ where $x_i \equiv m_{\wt e_i}^2/M_\alpha^2$; hereafter, we will consider just one flavor of Majorana, the DM candidate, and thus the index $``\alpha"$ will be implied. We have neglected the terms proportional to lepton masses,  and further assumed $m^2_h \ll m^2_{\wt e_i}, M^2$ in the last line}. $F_R$ can be obtained simply by exchanging $L\leftrightarrow R$. { We emphasize that to get Eq.~(\ref{FL}) which is not suppressed by small lepton masses we need both left- and right-handed scalar mediators,
which can be seen obviously from the fact it is proportional to $\mu$-parameter (See Eq.~(\ref{Lag}) and also the first
panel of Fig.~\ref{radiative}). The term with $\sin^2 2\theta$ comes from the contributions of ${\wt e_1}-{\wt e_1}$ and
${\wt e_2}-{\wt  e_2}$, while the term with $\cos^2 2 \theta$ comes from those of ${\wt e_1}-{\wt e_2}$ contributions in
the loop.
If we had a mediator with only one chirality, the chirality flip required in Eq.~(\ref{eq:h2mt}) would occur
only in external lepton lines. As a consequence the amplitude would be suppressed by small lepton masses and we could not
get sizable $h \to \mu\tau$ rate.}
In this paper, we follow the notations of three-point scalar function $C_0$ as in Ref.~\cite{Denner:1991kt}. The loop functions $G(x_1,x_2)=G(x_2,x_1)$ and $G(x_1)\equiv G(x_1,x_1)$ are defined in Eq.~(\ref{G:12}) and Eq.~(\ref{G:11}), respectively. 

As expected, in the decoupling limit with $\theta\ra 0$ (or $\pi/2$), the first term of $F_L$ is suppressed. 
In contrast, in the maximal mixing limit $\theta\ra \pi/4$, the second term is suppressed. Later, the former feature will be utilized 
to suppress LFV decay of charged leptons. 

\begin{figure}[htb]
\includegraphics[width=3.1in]{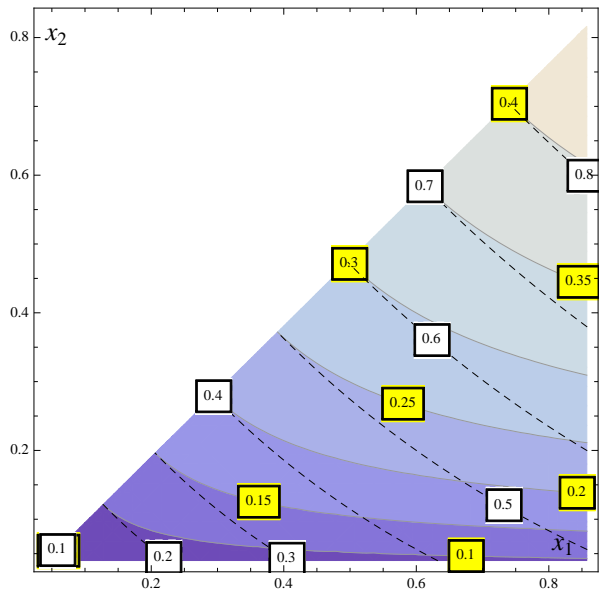}
\caption{Contour plots of loop functions $G(x^{{-1}}_1,x^{-1}_2)$ (solid lines) and $G(x^{-1}_1)+G(x^{-1}_2)$  (dashed lines). In the plot we use the variables $1/x_i$ instead of $x_i$; the same convention applies to other figures. }\label{GX12}
\end{figure}

The decay width of $h\ra\bar \ell_a \ell_b$ is calculated as
\begin{align}\label{}
\Gamma(h\ra\bar \ell_a \ell_b)=\f{m_h}{16\pi}\L |F_L|^2+|F_R|^2\R.
\end{align}  
For concreteness, we take $\ell_a=\tau$ and $\ell_b=\mu$ hereafter. In addition, for simplicity we consider only one chiral structure, i.e., 
setting $y_{L\tau}=y_{R\mu}=0$. It is easy to recover the corresponding contributions by the replacement $L\ra R$ and $R\ra L$ for 
all the later expressions. The implication of relaxation of this assumption will be commented when necessary. 
For reference, the branching ratio of $h\ra \bar\tau\mu$ is estimated in those two limits, the decoupling limit ($\theta\ra 0$):
\begin{align}\label{Br:h1}
{\rm Br}(h\ra\bar \tau \mu)=1.2\times10^{-2}\L\f{\mu}{\rm 5 TeV}\R^2\L\f{\rm 1 TeV}{M}\R^2 \L\f{G(x_1,x_2)}{0.2}\R^2  \L  \f{|y_{R\tau}y^*_{L\mu}|}{1}\R^2;
\end{align}  
and the maximal mixing limit ($\theta\ra \pi/4$):
\begin{align}\label{}
{\rm Br}(h\ra\bar \tau \mu)=1.2\times10^{-2}\L\f{\mu}{\rm 10 TeV}\R^2\L\f{\rm 1 TeV}{M}\R^2 \L\f{G(x_1)+G(x_2)}{0.4}\R^2  \L  \f{|y_{R\tau}y^*_{L\mu}|}{0.5}\R^2.
\end{align} 
The total decay width of Higgs boson has been taken to be 4 MeV. We show contour plots of $G(x_1,x_2)$ and $G(x_1)
\text{+} G(x_2)$ in Fig.~\ref{GX12}.

\subsection{ Induced CLFV $ \tau \ra\mu\gamma$}

The LFV decays of charged leptons are good probes to LFV. For example, the present experimental upper bound on 
Br$(\tau\ra \mu\gamma)$ is $4.4\times 10^{-8}$~\cite{tau:mugamma} at 90$\%$ C.L. and will 
be improved by one order of magnitude in the near future~\cite{tau:mugamma1}. 
The upper bound on CLFV { decay} of { muon} is even more stringent, Br$(\mu\ra e\gamma)<5.7\times10^{-13}$ at 90$\%$ C.L., from the current 
MEG result~\cite{CLFV:mu}. On the other hand, LFV Higgs decay is likely to induce CLFV decay (but not vice versa). 
Illustratively, the Feynman diagrams of the latter can be obtained simply by replacing the Higgs field with a photon leg in the charged loop of the  diagram for the former. As a schematic example, see the first and second panels of Fig.~\ref{radiative}. Since both processes share almost the same loops, a hierarchical ratio Br$(h\ra \tau\mu)/$Br($\tau\ra \mu\gamma$) as large as $10^{5}$ then raises doubt.

LFV decay of $\tau$ into $\mu+\gamma$ can be generically described by the following effective Hamiltonian:
\begin{align}
{\cal H}_{\rm eff} &= C_L \overline{\mu_L} \sigma^{\mu\nu} \tau_R F_{\mu\nu}
+C_R \overline{\mu_R} \sigma^{\mu\nu} \tau_L F_{\mu\nu}.
\end{align}
Different to significant chirality flip by virtue of the Higgs field in the loop of LFV Higgs decay process, here vector current conserves chirality. 
There are three other chirality violation sources to generate the Wilsonian coefficients $C_{L,R}$,
\begin{align}\label{CL}
C_L = &\frac{e}{32 \pi^2} 
\Bigg[ \Big\{ \f{m_\mu}{M^2} y_{R\mu} y^*_{R\tau} \L s_\theta^2 F_1(x_1)+c_\theta^2 F_1(x_2)\R
+\f{m_\tau}{M^2} y_{L\mu} y_{L\tau}^* \L c_\theta^2 F_1(x_1)+s_\theta^2 F_1(x_2)\R \Big\}
 \cr
 &~~~~~~~~~~-\f{1}{M}y_{L\mu} y^*_{R\tau} s_\theta c_\theta \L F_2(x_1)-F_2(x_2) \R
\Bigg].
\end{align}
The expression of $C_R$ can be obtained via $L\leftrightarrow R$. The loop functions $F_1(x)$ and $F_2(x)$ are defined as
\begin{align}
F_1(x) &= \frac{2+3x-6x^2+x^3 +6 x \log x}{12(1-x)^4}, \cr
F_2(x) &= \frac{-1+x^2 - 2 x \log x}{2 x (1-x)^2}.
\end{align}
According to the Hamiltonian, the decay width of $\tau \to \mu \gamma$ after summing over polarizations is calculated to be
\begin{align}
\Gamma(\tau \to \mu \gamma) = \frac{(m_\tau^2 - m_\mu^2)^3}{4 \pi m_\tau^3} \Bigg[ |C_L|^2+ |C_R|^2\Bigg].
\end{align}
In $C_L$, the first and the second terms do not require the simultaneous presence of $y_L$ and $y_R$ because chirality
flip comes from the external lines, i.e., the Dirac mass term of lepton. But they require LFV through the same chirality
of slepton. These contributions are generically subdominant, compared to the third term, given a large $M$ and as well
democratic type Yukawa coupling, {\it i.e.}, $y_L\sim y_R$. Besides, a sizable mixing angle between $\phi_\ell^+$ and
$\phi_e$ is needed. This means that, not only flavor violation but also chirality violation are provided by the
sleptons, as is well understood from the second panel of  Fig.~\ref{radiative}.

{ We argue that the $h \to \tau \mu$ rate can be enhanced while suppressing $\tau \to \mu \gamma$. (See Sec.~\ref{sec:HLFV} for more details.)
One obvious mechanism is to use heavy $\phi_e$, which naturally leads to small mixing angle $\theta$.
In this case the $\tau \ra \mu\gamma$ diagram has one more $\phi_e$ propagator compared with the $h \to \tau\mu$ diagram
as shown in the first two diagrams in Fig.~\ref{radiative}, suppressing the former compared to the latter.
}

\subsection{ Hints in $h\ra\gamma\gamma$}

Since LFV Higgs decay heavily depends on the charged scalar mixing term,  $h\ra\gamma\gamma$ inevitably receives a sizable contribution. Under the assumption that other Higgs decay modes are not affected, which is a very natural assumption, we get the modification to $h\ra\gamma\gamma$ from the $\wt e_1-$loop~\cite{hFF}~\footnote{In the following
  analysis we decouple the $\wt e_2-$loop by assuming much heavier $\wt e_2$, otherwise the contributions from the $\wt
  e_1-$ and $\wt e_2-$loop show substantial cancellation: { $\delta c_\gamma\approx-\f{v\mu\sin2\theta  }{48\sqrt{2}}
    \L\f{1}{m_{\wt e_1}^2}-\f{1}{m_{\wt e_2}^2}\R$}. Then the bound becomes weaker.\label{degen}}, 
\begin{align}
c_\gamma=c_{\rm SM,\gamma}+\delta c_\gamma\approx-0.81 {-\f{1}{24} \f{v\mu\sin2\theta  }{2\sqrt{2}m_{\wt e_1}^2}}
\end{align}
Here $c_\gamma$ denotes the reduced coupling of the dimension-five operator  for coupling between Higgs and photons, $c_\gamma\f{\alpha}{\pi v} hF_{\mu\nu}F^{\mu\nu}$. The sign of $\mu$ is indeterminate, so one can make $r_\gamma$ close to the SM value either by requiring a small $\delta c_\gamma\ll1$ or $\delta c_\gamma\sim +1.62$, which flips the sign of $c_\gamma$ relative to the SM one. To be more specific, we refer to a recent study~\cite{Bernon:2014vta}, 
from which we know that at 68.3$\%$ C.L. there are two allowed regions:
\begin{align}
-0.05\lesssim \delta c_\gamma/c_{\rm SM,\gamma}\lesssim0.20,\quad -2.20\lesssim \delta c_\gamma/c_{\rm SM,\gamma}\lesssim -1.95.
\end{align}
Feeding { these results} back to the slepton sector we get the following constraints:
\begin{itemize}
\item In the first region, one gets the bounds:
{
\begin{align}\label{mu:small}
 -1.0\times \L\f{m_{\wt e_1}}{300\rm GeV}\R^2{\rm TeV}\lesssim \mu\sin2\theta \lesssim   4.0\times \L\f{m_{\wt e_1}}{300\rm\GeV}\R^2\rm {TeV}.
\end{align}}
As one can see, as long as $\wt e_1$ mass is at least a few hundred GeVs, the Higgs diphoton rate in the decoupling limit can be easily suppressed below the upper bound. But it is not that easy to reconcile Br$(h\ra\tau\mu)$ and  Br$(h\ra \gamma\gamma)$ in the maximal mixing limit. The $\wt e_1$ should be sufficiently heavy, or it should have roughly equal mass with $\wt e_2$ in the light of footnote~\ref{degen}. 

\item The second region allows for the scenarios with a huge $\mu$ along with a lighter $m_{\wt e_1}$. In this way of reconciling Br$(h\ra\tau\mu)$ and  Br$(h\ra \gamma\gamma)$, it (asides from determining the sign of $\mu$) actually helps to eliminate one of the three parameters in the slepton sector: 
\begin{align}
\mu\sin2\theta\approx 40.2\times \L\f{m_{\wt e_1}}{300\rm\GeV}\R^2\L\f{\delta c_\gamma}{-2.0c_{\rm SM,\gamma}}\R\rm {TeV}.
\end{align}
A TeV scale $m_{\wt e_1}$ will blow up $\mu$, thus disfavored. By the way, a too large $\mu/m_{\wt e_1}\gg 10$ may also change Higgs self-coupling too much.

\end{itemize}
In summary, Higgs diphoton does not give a severe constraint. But it is interesting to see that possibly the rate can be related to the large LFV Higgs decay.

\subsection{Natural ways to get large $h\ra \tau \mu$}
\label{sec:HLFV}

We have collected all the necessary formulas to calculate Br$(h\ra \tau \mu)$ { under the constraints such as Br$(\tau \to \mu\gamma)$}. In this subsection we show how Br$(h\ra \tau \mu)\sim10^{-2}$ { can be} realized. For that, it is convenient to study the ratio
$R_\tau\equiv{ {\rm Br}(h\ra \tau\mu)}/{{\rm Br}(\tau\ra \mu\gamma)}$. 
{ To explain the central value of the $h \to \tau\mu$ signal, $B(h \to \tau\mu) \approx 0.85 \%$, with the contraint
$B(\tau \to \mu\gamma) <4.4 \times 10^{-8}$, we need $R_\tau \gtrsim 2 \times 10^5$.} In the decoupling limit of the scalar system, { $R_\tau$} can be illustratively parameterized as
\begin{align}\label{Rtau}
R_\tau\approx   { 2.8}\times 10^{5}\L\f{\mu}{10\rm\,TeV}\R^2 \L\f{0.1}{\sin\theta}\R^2 \L \f{G(x_1,x_2)/(F_2(x_2)-F_2(x_1))}{20}\R^2. 
\end{align}
We have made the approximation that Eq.~(\ref{FL}) and Eq.~(\ref{CL}) are dominated by the second and third terms, respectively. 
In this approximation, $R_\tau$ is independent { of} (or insensitive to) the following parameters: 
(I) DM mass $M$; (II) the Yukawa couplings; (III) to some degree,  also $\mu$. To see the last point, from Eq.~(\ref{theta}) one may have $1/\sin2\theta\approx m_{\wt e_2}^2/\sqrt{2}\mu\nu$ and consequently $\mu^2$ is cancelled. This conclusion holds for a well asymmetric scalar system like $m_{\phi_\ell}^2\gg m_{\phi_e}^2,  2\mu^2v^2$, which guarantees decoupling scalars as desired. If instead the scalar sector is in the maximal mixing limit and thus Eq.~(\ref{FL}) is dominated by the first term, we have the estimation
\begin{align}\label{Rtau:mix}
R_\tau\approx  { 2.8}\times 10^{5}\L\f{\mu}{10\rm\,TeV}\R^2 \L \f{(G(x_1)+G(x_2))/(F_2(x_2)-F_2(x_1))}{200}\R^2. 
\end{align}
In the absence of enhancement from (the inverse of) small mixing, one needs a huge $\mu$ at least 10 TeV and at the same time a very large ratio $(G(x_1)+G(x_2))/(F_2(x_2)-F_2(x_1))\sim{\cal O}(100)$. While in the previous case it is moderate.
That large ratio may incur a significant fine-tuning. In order to lift the ratio, one needs
cancelation~\footnote{Ref.~\cite{Dorsner:2015mja} also considered cancelations in $\tau\ra\mu\gamma$ via introducing
  some extra contributions to cancel the contribution induced by $h\ra\tau\mu$. In our model this is kind of cancelation
  happens within well expectation.} between $F_2(x_2)$ and $F_2(x_1)$. Obviously, if $x_1\approx x_2$, cancelation
happens.~\footnote{Cancelation also happens for $x_1\neq x_2$. In particular, for a (at least) mild mass hierarchy between $m_{\wt e_1}^2$ and $m^2_{\wt e_2}$, cancelation approximately determines $M$: $M\simeq\f{m_{\wt e_1}m_{\wt e_2}}{\sqrt{3m_{\wt e_1}^2+m_{\wt e_2}^2}}.$}

\begin{figure}[htb]
\includegraphics[width=3.0in]{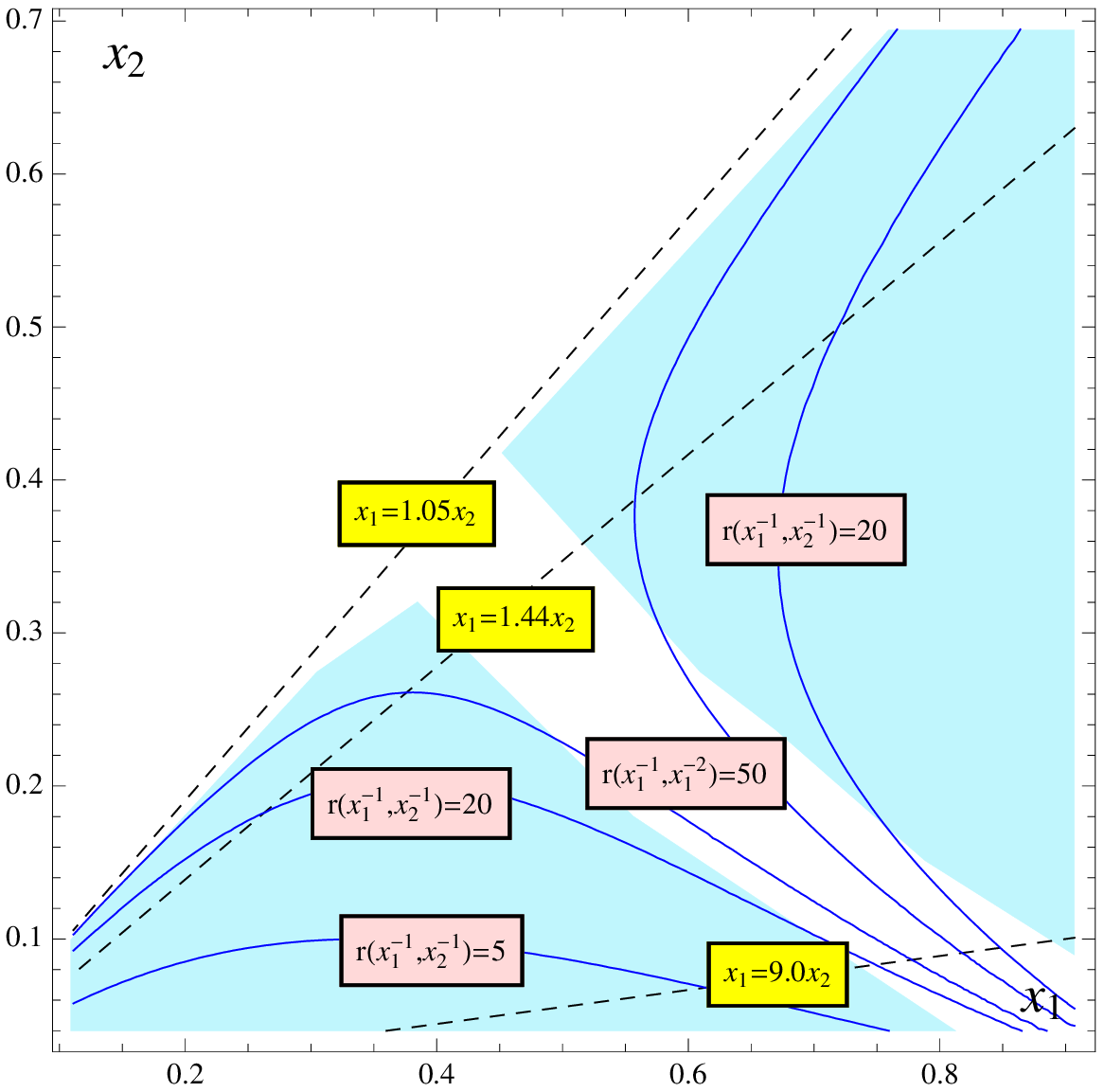}~~
\includegraphics[width=3.0in]{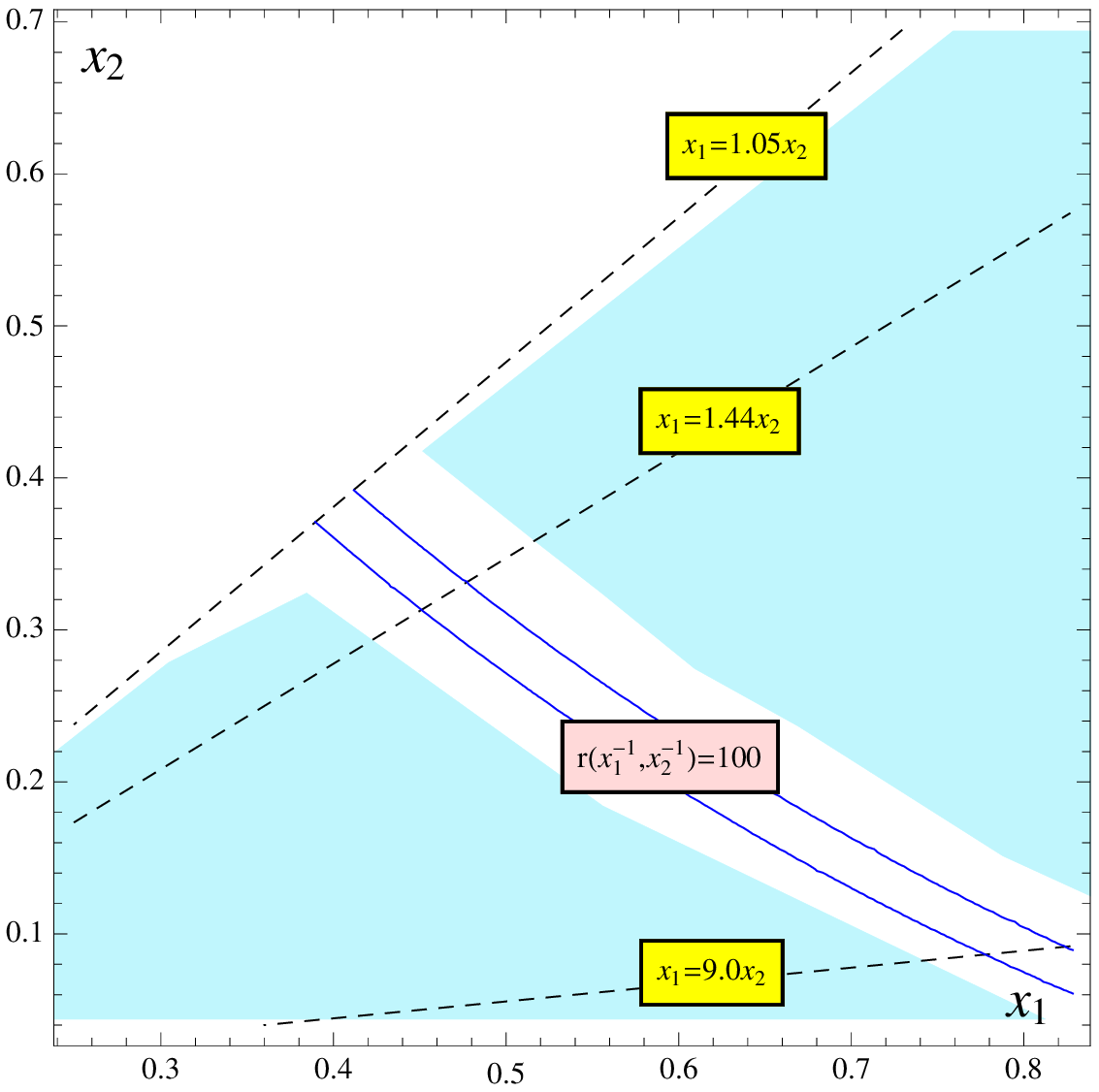}
\caption{Contour plots of loop functions ratio $r(x^{-1}_1,x^{-1}_2)$ (blue lines), which is $G(x^{-1}_1,x^{-1}_2)/(F_2(x^{-1}_2)-F_2(x^{-1}_1)$ in the decoupling limit (left) and $(G(x^{-1}_1)+G(x^{-1}_2))/(F_2(x^{-1}_2)-F_2(x^{-1}_1)$ in the maximal mixing limit (right panel). Regions with fine-tuning better than $5\%$ are shaded. Besides, we label three selected ratios of the masses of two charged scalars (dashed lines).}\label{FT:r}
\end{figure}
Regarding $F_2(x_2)-F_2(x_1)$ as a function of three fundamental variables $\mu_i=(m_{\wt e_1},\,m_{\wt e_2},\,M)$, we can measure fine-tuning using the quantity
\begin{align}
\Delta=\max\{|\Delta_i|\}|_{i=1,2,3}~~{\rm with }~~\Delta_i\equiv \f{\partial \log (F_2(x_2)-F_2(x_1))}{\partial\log \mu_i}.
\end{align}
Explicitly, $\Delta_i=2\L{c_{i2}x_2F'_2(x_2)-c_{i1}x_1F'_2(x_1)}\R/{(F_2(x_2)-F_2(x_1))}$ with $c_{11,12}=(1,0)$, $c_{21,22}=(0,1)$ and { $c_{31,32}=(-1,-1)$}.

Let us denote the ratio of loop functions in Eq.~(\ref{Rtau}) and Eq.~(\ref{Rtau:mix}) as $r(x_1,x_2)$. In Fig.~\ref{FT:r}, we plot the distributions of $r(x^{-1}_1,x^{-1}_2)$ and fine-tuning $\Delta$ on the $x_1-x_2$ plane. The left  and the right panel are for the decoupling and the maximal mixing scenarios, 
respectively. The shaded regions have degree of fine-tuning less than $5\%$, which is referred as the lower bound for naturalness in this article. It is seen that the decoupling scenario can provide $r(x^{-1}_1,x^{-1}_2)\sim{\cal O}(10)$ barely incurring fine-tuning; in contrast, the maximal mixing scenario, which needs  $r(x^{-1}_1,x^{-1}_2)\gtrsim 100$, typically incurs fine-tuning worse than $5\%$. But the cancelation via degenerate $\wt e_1$ and $\wt e_2$ still opens a narrow region around the point $x^{-1}_1\simeq x^{-1}_2\sim 0.6$ or closely alone the line $x_1=x_2$, which without a particular UV reason is not of much interest. In what follows we will focus on these two kinds of natural regions.

Let us consider the decoupling scenario. We make several observations that are helpful to trace back to the patterns of scalar mass squared matrix. 
\begin{enumerate}
\item If both $x^{-1}_{1,2}\ll1$, we need significant degeneracy between two scalars, see the left-bottom corner of the left panel of Fig.~\ref{FT:r}. Since we are chasing the decoupling limit confronting a large slepton mixing term with $\mu\sim{\cal O}(10)$ TeV, this means large and degenerate scalar mass terms $m_{\phi_\ell}^2\sim m_{\phi_e}^2\gg {\cal O}(1)\rm TeV^2$. It results in a heavy spectrum typically having multi-TeV sleptons, see the left panel of Fig.~\ref{mixing}.
\item There is a hierarchy $x^{-1}_2\lesssim {\cal O}(0.1)x^{-1}_1$, keeping $x^{-1}_1$ close to 1. It requires an asymmetric scalar system, e.g., $m_{\phi_\ell}^2\gg m_{\phi_e}^2\sim {\cal O}(1) \rm TeV^2$, the most favored pattern to decouple $\phi_\ell$ and $\phi_e$ with a large mixing term. 
\item $x_2^{-1}\lesssim x_1^{-1}$, both not far from 1. This is in the bulk space without special requirements. Even for a smaller $\mu$ near the TeV scale, one is still able to produce such a case readily, yielding a lighter spectrum inducing DM.
\end{enumerate} 
In summary, there is a wide parameter space for the decoupling scenario. In practice, in some situations the mixing angle is supposed to be moderately small rather than very small,

\section{Leptophilic DM: relic density \& direct detection}
\label{sec:dm}

The DM candidate $N$~\footnote{In our model, in principle DM can be either the neutral component of Higgs doublet
  $\phi_\ell$ or the Majorana fermion  { $N$}. But only the fermonic DM could be a natural leptophilic DM.}  is a singlet Majorana fermion with $t-$channel mediators, and its phenomenologies in some simplified cases have been investigated compressively in Ref.~\cite{Garny:2015wea}. But our case turns out to be significantly different, due to the appearance of both $\phi_\ell$ and $\phi_e$ mediators. In this section we will focus on two main differences, annihilation and direct detection of DM.

\subsection{Annihilation: $s-$wave versus $p-$wave}

The first difference comes from DM annihilating. The Majorana DM $N$ annihilates into leptons through the interactions given in Eq.~(\ref{model:com}). They proceed with $\wt e_i$ exchanging in the $t-$ and $u-$channel. We can calculate the cross section expanded in terms of DM relative velocity $v_r\equiv 2\sqrt{1-4M^2/s}$ in the center-of-mass (CM) frame: $\sigma v_r\approx a+b v_r^2$ with the $s-$ and $p-$wave coefficients respectively given by
\begin{align}\label{}
a=&\f{1}{16\pi M^2}\f{1}{(1+x_i)^2}\L |\ld_{ia}^L\ld_{ib}^R|^2+|\ld_{ia}^R\ld_{ib}^L|^2\R,\\
b=&\f{1}{96\pi M^2}\f{1}{(1+x_i)^4}\left[ 2|\ld_{ia}^L|^2|\ld_{ib}^L|^2 (1+x_i)- |\ld_{ia}^L|^2|\ld_{ib}^R|^2 \L 1+4x_i-3x_i^{2}\R +(L\leftrightarrow R) \right].
\end{align} 
The inclusive annihilation rate should sum over the family index $a$ and $b$.  As a check, when the model goes to the chiral limit considered previously~\cite{Garny:2015wea}, e.g., $\ld_{i}^R~ ({\rm or} ~y_{L})\ra0$, we recover the well known result: $a=0$ (up to contributions suppressed by lepton masses). Then, DM must annihilate away mainly via $p-$wave, whose coefficient takes a form of
\begin{align}
b\ra &\f{1}{48\pi M^2}\f{1+x_i}{(1+x_i)^4}|\ld_{ia}^L\ld_{ib}^L|^2.
\end{align} 
It is not suppressed by small mixing. For instance $\theta\ra0$, it still receives a contribution from $|\ld_{2a}^L\ld_{2b}^L|^2\ra| y_{Ra} y_{Rb}|^2$. With them, the relic density can be calculated via the well-known formula~\cite{coann}
\begin{align}
\Omega h^2\approx \f{0.88\times 10^{-10}x_f\rm GeV^{-2}}{g_*^{1/2} (a+3b/x_f)}. 
\end{align}
At the freeze-out epoch $x_f=M/T_f\sim 20$, the effective degree of freedom $g_*\sim 100$. If we demand the Yukawa coupling constants~$\lesssim {\cal O}(1)$, in order to maintain perturbativity of the model up to a very high scale, then both DM and mediators should around the weak scale. This is a strong requirement and yields deep implication to direct detection.

But here the $s-$wave may be sufficient to reduce the DM number density, even facing the stringent CLFV constraint and
at the same time satisfying the tentative LFV Higgs decay. It is seen that the $s-$wave coefficient is directly correlated with CLFV decay width $\Gamma(\ell_a\ra \ell_b \gamma)$, see Eq.~(\ref{CL}). To be more specific, we write $a$ in terms of others
{
\begin{align}
 a\approx &\f{1}{64\pi M^2}\sin^22\theta\L  |y_{La}y_{Rb}|^2+|y_{Ra}y_{Lb}|^2\R\sum_i\f{1}{(1+x_i)^2}.
\end{align} 
}
It may reach the typical cross section of thermal DM, 1 pb. To see this, we parameterize the order of magnitude of $a$ as the following:
\begin{align}
 a\simeq &0.8\times \L\f{400\rm\,GeV}{M}\R^2\L\f{\sin^2\theta}{0.1}\R\f{\L  |y_{La}y_{Rb}|^2+|y_{Ra}y_{Lb}|^2\R}{1.0}\rm pb.
\end{align} 
We have taken { $1/(1+x_1)^2\approx0.15$}. Therefore, again a weak scale DM along with (at least one) weak scale mediator can lead to correct relic density via $s-$wave annihilation as long as the mixing angle is not highly suppressed. 

Although the $s-$wave annihilation readily works for flavors like $a=b$ which does not violate lepton flavor, it fails for the case under consideration $a=3,\,b=2$ or inverse. Let us show it in the decoupling scenario. With the aid of  Eq.~(\ref{Br:h1}) and Eq.~(\ref{Rtau}) we can express $a$ as (aside from the propagator factor)
\begin{align}
 a\simeq &2.1\times \L\f{{\rm Br}(h\ra \tau \mu)}{10^{-2}}\R \L\f{3\times10^{5}}{R_\tau}\R \L\f{10^{-4}}{F_2({x_2})-F_2({x_1})}\R^2 \rm pb.
 \end{align} 
But that small value of $F_2({x_2})-F_2({x_1})$ either incurs large fine-tuning or should follow closely the line $x^{-1}_1\simeq x^{-1}_2\ll0.1$. The latter leads to additional suppression $\sim 1/x_1^2$ (it has been fixed to be 0.15 in the above estimation). Similarly, the maximal mixing scenario fails either.

We make a comment on the coannihilation effect~\cite{coann}. Despite of not a focus here, it has two interesting points. First, mass degeneracy between $\wt e_1$ and $M$ is well consistent with the suppression of Br$(\tau\ra\mu\gamma)$, which is made small by the cancellation mechanism with $x_1\neq x_2$. For a strong mass hierarchy case $m_{\wt e_2}^2\gg m_{\wt e_1}^2$, from footnote 4 we have $M\approx m_{\wt e_1}$. Second, by virtue of a large $\mu-$term, the effective cross section of coannihilation is enhanced by the process $\wt e_1^+\wt e_1^-\ra hh$ with $\wt e_1$ in the $t-$channel:
\begin{align}
\sigma_{hh} v&\approx \f{1}{64\pi}\f{1}{m_{\wt e_1}^2}\L\f{\mu^2/2}{m_{\wt e_1}^2}\R^2\cr
&=1.2\times 10^{-5}\L\f{1\rm\,TeV }{m_{\wt e_1}}\R^2\L \f{\mu/m_{\wt e_1}}{10}\R^4\rm GeV^{-2}.
\end{align} 
We have worked in the maximal mixing $\sin2\theta\approx1$. So, once $\mu\sim{\cal O}(10\rm TeV)$, the enhancing factor still scales as $\L\mu/m_{\wt e_1}\R^4\sim {\cal O}(10^4)$ even for a TeV scale $m_{\wt e_1}$.

\subsection{On (in)direct searches for the leptophilic DM}

We have shown that DM can gain correct relic density readily. And DM mass should be around the weak scale so as to avoid large Yukawa couplings. In this subsection we move to the second difference, direct detection. As a leptophilic DM, DM-nucleon scattering is absent at tree level, but could be generated by radiative corrections. There are two types of corrections leading to DM-nucleon scattering, one mediated by photon and the other Higgs boson, respectively. In particular, the second type, which is absent in the previous setup, benefits from $\mu-$enhancement and can potentially overcome the loop suppression.

The second type is the usual dimension-four operator which comes from the vertex correction on $h\bar N N$, absent at tree level but generated after EWSB. In the DM direct detection, typically the transferring momentum $Q^2$ is very small compared with the other mass scales in the charged particles in the loops, so that
\begin{align}
{\cal O}_h=\ld_{hN}(0)h\bar N N,
\end{align} 
where $h$ is treated off-shell with invariant mass $Q^2\ll m_h^2$. The effective coupling at zero momentum transfer  is expressed as 
\begin{align}\label{}
\ld_{hN}(0)&\approx \sin\theta\f{|y_{La}|^2+|y_{Ra}|^2}{32\pi^2}\f{\mu}{\sqrt{2}M}\left[     B_0(p_1-p_2)_{a1}- B_0(p_1-p_2)_{a2}-2 B_0(p_1)_{11} +2 B_0(p_1)_{12}
 \right.& \nonumber \\
&\left.
-C_0(-p_2,p_1-p_2)_{a12} \L m_{\wt e_1}^2+m_{\wt e_2}^2-2M^2\R+2C_0(-p_2,p_1-p_2)_{a11}\L m_{\wt e_1}^2-M^2\R   \right].
\end{align}  
Using the kinematics and the approximations of two- and three-point scalar functions in Appendix~\ref{approx}, we can further simplify it into 
\begin{align}\label{}
\ld_{hN}(0)&\approx \sin\theta\f{|y_{La}|^2+|y_{Ra}|^2}{32\pi^2}\f{\mu}{\sqrt{2}M}\left[ 2+ {(x_1-1) \log(1-x^{-1}_1)} -{(x_2-1) \log(1-x^{-1}_2)}
 \right.& \nonumber \\
&\left.
-\f{1-x^{-1}_1}{1-x^{-1}_2}\log\f{x_2}{x_1}-{\cal G}(x_1,x_2)\L{x_2}+{x_1}-2\R+2{\cal G}(x_1,x_1)\L{x_1}-1\R \right],
\end{align}
with ${\cal G}(x_1,x_2)$ seen in Eq.~(\ref{g12}). Note that $x_1\simeq x_2$ shows cancellation and thus larger $\ld_{hN}(0)$ dwells on the region with $x_1$ at least modularity larger than $x_2$. 
\begin{figure}[htb]
\includegraphics[width=3.0in]{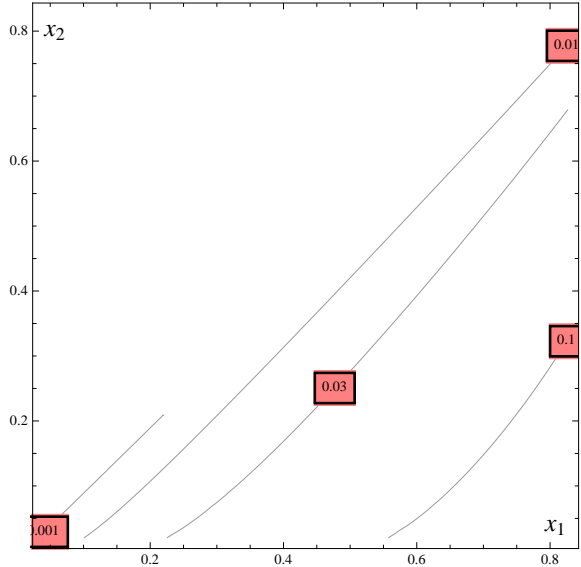}
\caption{Contour plots of $\ld_{hN}(0)$ (solid lines) in the maximal scenario, setting $\mu=10$ TeV, $M=200$ GeV and either $y_L=1$ or $y_R=1$; Again, the variables in this plot are $1/x_i$ not $x_i$.}\label{}
\end{figure}

The Higgs mediated DM-nucleon scattering has a spin-independent cross section $\sigma^p_{\rm SI}=4m_p^2f_p^2/\pi$ with $f_p$ given by
\begin{align}
f_p=\f{\ld_{hN}(0)}{m_h^2}m_p\left[   \sum_{q=u,d,s}y_q \f{f_{T_q}^{(p)}} {m_q}+\f{2}{27}{f_{T_G}^{(p)}}\sum_{q=c,b,t} \f{y_q} {m_q} \right].
\end{align} 
In this paper we take $f_{T_u}^{(p)}=0.020$, $f_{T_d}^{(p)}=0.026$, $f_{T_s}^{(p)}=0.118$ and $f_{T_G}^{(p)}=0.840$~\cite{X. Gao} (For more discussions about the calculation and uncertainties of these values, see Refs.~\cite{Alarcon:2011zs}) and we then get the estimation $\sigma^p_{\rm SI}\approx 4.0\times 10^{-8}\L\ld_{hN}(0)/0.1\R^2 \rm pb$, a value near the sensitivity of the current LUX. In the bulk parameter space, $\ld_{hN}(0)\lesssim {\cal O}(0.01)$:
\begin{align}
\ld_{hN}(0)&\approx 0.01\times\L\f{\sin\theta}{0.2}\R\L\f{|y_{La}|^2+|y_{Ra}|^2}{1}\R \L\f{\mu}{5\rm TeV}\R\L\f{0.3\rm TeV}{M}\R.
\end{align}
The decoupling scenario is hard to be probed, but the maximal mixing scenario, which badly needs a very large $\mu$, has a good prospect. We choose a benchmark case which is directly related with $h\ra \tau\mu$.

Photon-mediated scattering becomes important for lighter mediators. Since our DM is a Majorana fermion, the leading order operator for DM-nucleon coupling  is the dimension-six anapole operator~\cite{anpole}:
\begin{align}
{\cal O}_A={\cal A}\bar N\gamma^\mu \gamma^5 N \partial^\nu F_{\mu\nu},
\end{align} 
The ${\cal A}$ can be obtained by integrating out loopy particles step by step~\cite{Agrawal} or via direct calculation of the loops~\cite{Kopp}:
\begin{align}
{\cal A}\approx-\f{e\L|\ld_{ia}^L|^2+|\ld_{ia}^R|^2\R}{192\pi^2 M^2}\left(-3\log({x_i \epsilon_a}) -\f{x_i+3}{1-x_i}{\log \f{x_i^{-1}-1}{\sqrt{\epsilon_a}}}  \right),
\end{align} 
with $\epsilon_a=m_{\ell_a}^2/M^2$. The expression is valid for the heavy leptons with $m_{\mu,\tau}^2\gg |Q|^2$. It is seen that ${\cal A}$ is insensitive to the $\mu-$term and the mixing angle. For $M=100$ GeV, it is estimated that ${\cal A}/\L|\ld_{ia}^L|^2+|\ld_{ia}^R|^2\R\sim {\cal O}(10^{-7})\rm GeV^{-2}$. The resulting scattering rate is at least four orders of magnitude weaker than the current LUX sensitivity~\cite{Kopp}.

\section{Conclusion}
\label{sec:concl}

In SM, lepton flavor is accidentally conserved but on the other hand LFV is an established fact. So it is of importance to search for LFV processes such as LFV Higgs decay in the LHC era. It is a good probe to new physics. But LFV Higgs decay is negligible and undetectable in most new physics models for addressing neutrino masses. In this paper we study a type of new physics that could lead to large Higgs LFV decay, i.e., lepton-flavored dark matter specified by the particle property of DM (a Majorana fermion) and DM-SM mediators (scalar leptons). Different than other similar setups, here we introduce both the left-handed and the right-handed scalar leptons. They allow for large LFV in Higgs decay and thus may explain the tentative Br$(h\ra\tau\mu)\sim1\%$. In particular, we find that the stringent bound from $\tau\ra\mu\gamma$ can be naturally avoided especially in the decoupling limit of slepton sector. Aspects of relic density and radiative direct detection of the leptonic DM are also investigated.

There are several open questions that deserve future investigation. First, as mentioned in the text, neutrino masses and mixings can be radiatively generated because all the core of the Ernest Ma's model is already incorporated in our model. Even restricted to one RHN, i.e., the Majorana DM, we are able to generate realistic neutrino mixings after introducing a couple of scalar lepton doublets $\phi_{l,i}$. Second, in this article we merely discuss LFV in the first and second family of leptons, and such kind of discussions are easily generalized to other families, which is of particular interest when correlated with neutrino phenomenologies. However, it is not easy to reconcile the tiny neutrino mass scale with a large LFV  Higgs decay like Br$(h\ra\tau\mu)\sim1\%$, because the former basically requires somewhat smaller Yukawa couplings $~{\cal O}(0.01)$. Of course, if we work on very light DM like below the GeV even MeV scale, maybe there still stands a chance.

\section*{Acknowledgement}
We thank Pyungwon Ko very much for valuable discussions and reading the manuscript carefully.
This work is supported in part by  National Research Foundation of Korea (NRF) Research Grant 
NRF-2015R1A2A1A05001869 (SB).
 
\appendix

\section{Two- and three-points scalar functions and their limits}\label{approx}

In this appendix we present the technical details used in this paper. The scalar three point function is defined as~\cite{oneloop,Denner:1991kt}. 
\begin{align}
C_0(p_1,p_2,m_0,m_1,m_2)
=&{\f{(2\pi\mu)^{4-d}}{i\pi^2}\int d^dk\f{1}{\L k^2-m_0^2\R\L(k+p_1)^2-m_1^2\R\L(k+p_2)^2-m_2^2\R} }\cr
=&-\int_{0}^{1}dx\int_0^x  dy  \f{1}{ ax^2+by^2+c xy+dx+e y+f-i\epsilon }
\end{align} 
with 
\begin{align}\label{C0:1}
a=&(p_2-p_1)^2,\quad b= p_1^2, \quad c=p_2^2-p_1^2-(p_2-p_1)^2,\cr
d=&m_1^2-m_2^2-(p_2-p_1)^2,\quad e= m_0^2-m_1^2+(p_2-p_1)^2-p_2^2,\quad f=m_2^2.
\end{align} 
When $p_1^2=p_2^2$, obviously we have $C_0(p_1,p_2,m_0,m_1,m_2)=C_0(p_1,p_2,m_0,m_2,m_1)$. If the invariant masses of
the external momentums $p_{1,2}^2,\,(p_2-p_1)^2$ are far lighter than the mass scales of the particles
{ in the loop}, $m_{0,1,2}^2$, one can approximate $C_0(p_1,p_2,m_0,m_1,m_2)$ to be
\begin{align}\label{G:12}
C_0(m_0,m_1,m_2)= -\f{1}{m_0^2}G(r_1,r_2)\quad {\rm with }\quad { G(r_1,r_2) \equiv \f{1}{r_1-r_2}\L \f{r_1\log  r_1}{r_1-1} -\f{r_2\log r_2}{r_2-1}\R,}
\end{align} 
with $r_i\equiv m_i^2/m_0^2$. Note that $G(r_1,r_2)$ is symmetric under { interchanging} $r_1$ and $r_2$. There are two particular limits that are helpful in analyzing the radiative decays of Higgs boson.
\begin{description}
\item[$ r_2=r_1$ ] For this single propagator case one has
\begin{align}\label{G:11}
C_0(m_0,m_1,m_1)=-\f{1}{m_0^2}G(r_1)
{ \equiv - \f{1}{m_0^2}\f{r_1-1-\log r_1}{\L r_1-1\R^2}}.
\end{align} 
If further $r_1$ goes to 1, it slides to ${1}/{2m_0^2}$. But for very heavy $m_1$ it decouples as $1/m_1^2$. 
\item[ $r_2\gg r_1\ra1$] For the asymmetric propagators like this, we have the simple approximation 
\begin{align}
{ C_0(m_0,m_1,m_2)\doteq { -}\f{1}{m_0^2}\f{1}{(r_2-1)^2} \L 1-r_2+r_2\log r_2\R\approx { -} \f{1}{m_2^2} \L \log r_2-1\R,}
\end{align} 
Due to the logarithmic factor, it decouples slower than the previous case. 
\end{description}
\begin{figure}[htb]
\includegraphics[width=3.0in]{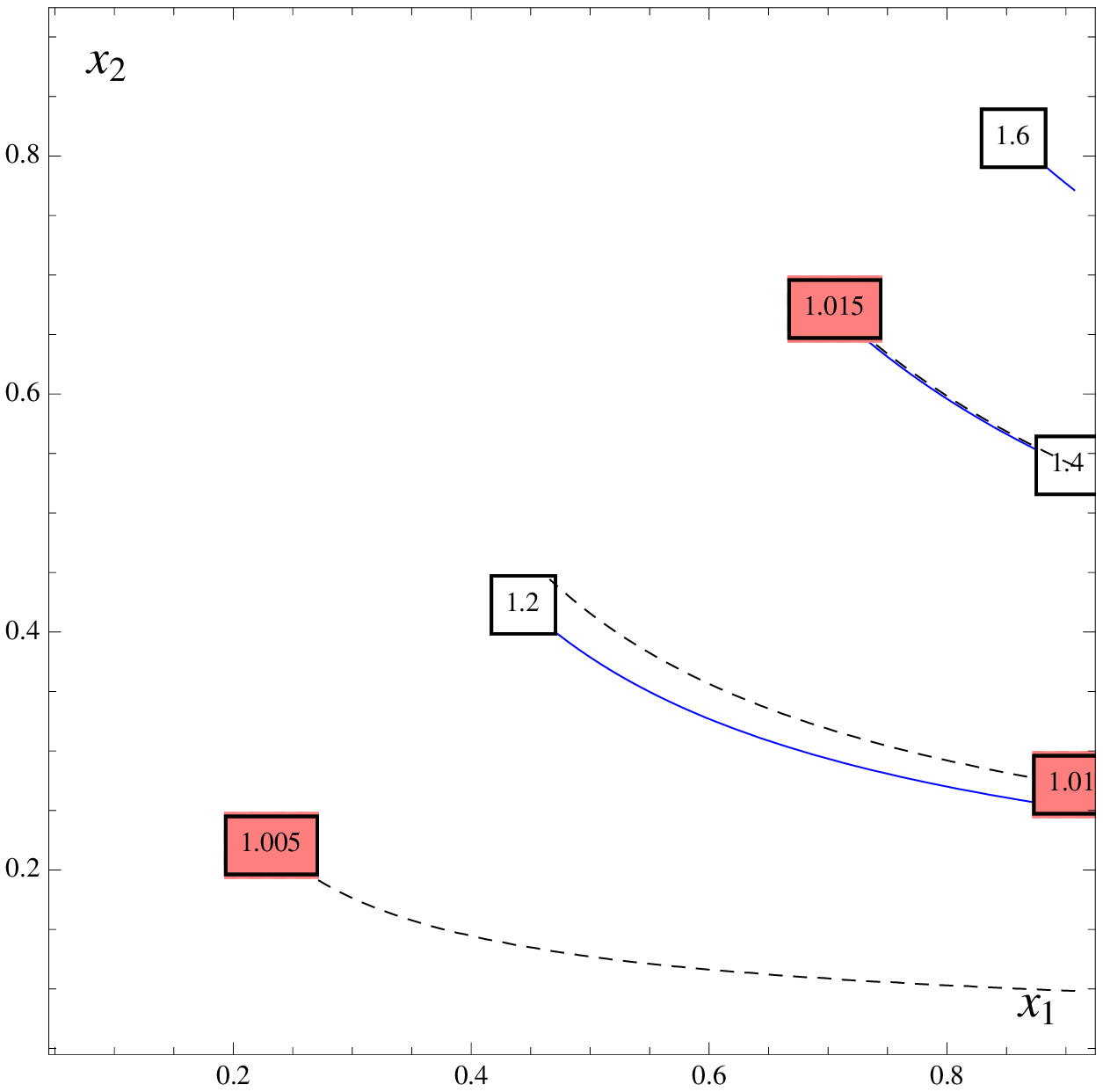}
\includegraphics[width=3.0in]{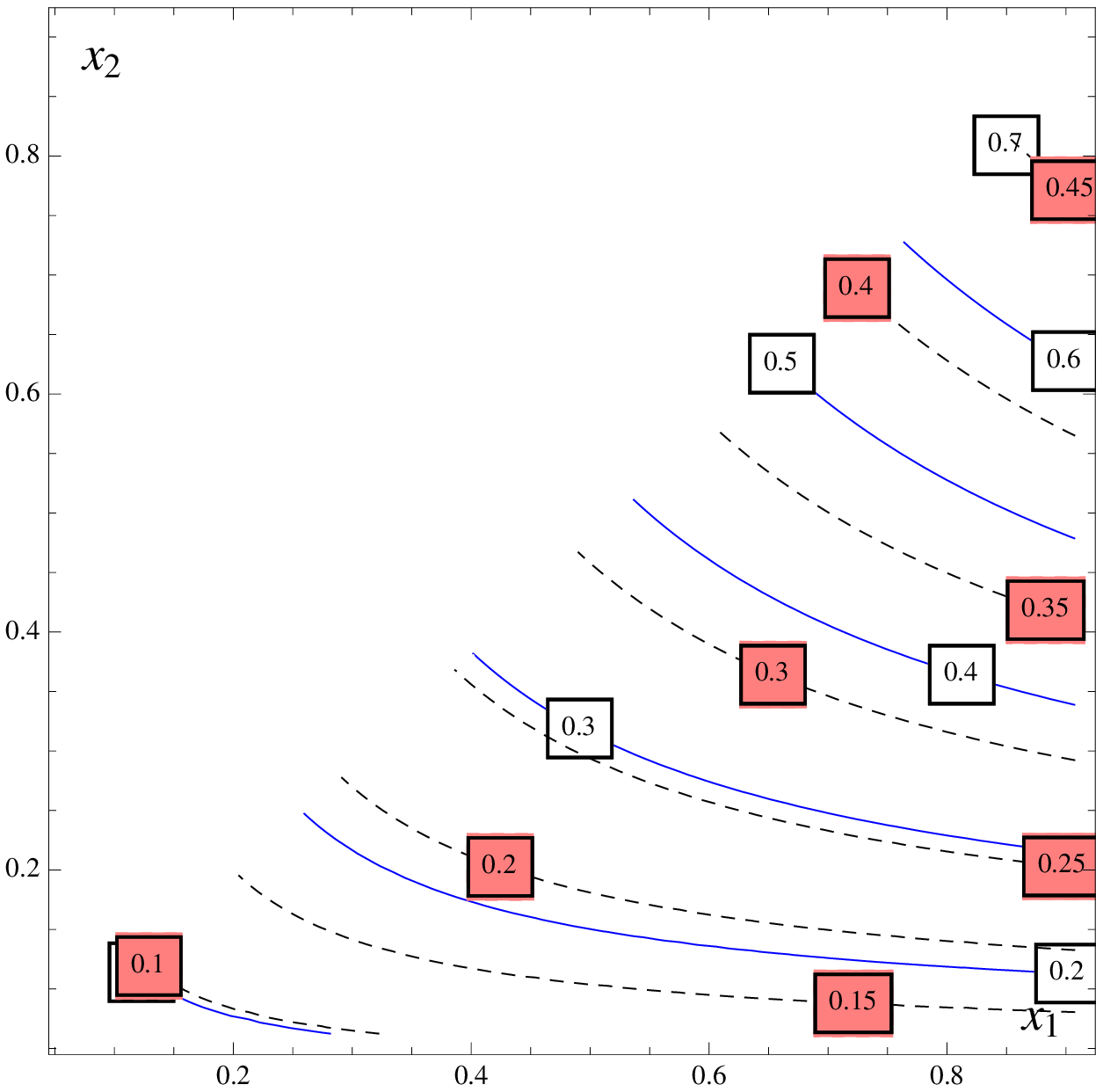}
\caption{Left: Test of approximation; right: Distributions of loop integrals. In both panels the dashed lines are for $(p_1-p_2)^2=0.25m_0^2$ and solid lines for $(p_1-p_2)^2=4m_0^2$; $x_i=m_0^2/m_i^2.$}\label{}
\end{figure}

The scalar two-point function is defined as 
\begin{align}\label{B0}
B_0(p_1,m_0,m_1)=\int d^dk \f{1}{(k^2-m_0^2)((k-p_1)^2-m_1^2)},
\end{align} 
which satisfies the relations $B_0(p_1,m_0,m_1)=B_0(p_1,m_1,m_0)$ and $B_0(p_1,m_0,m_1)=B_0(p_2,m_0,m_1)$ for $p_1^2=p_2^2$. Actually, it has an explicit expression (up to ${\cal O}(\epsilon)$)
\begin{align}
B_0(p_1,m_0,m_1)=\Delta+2-\log\f{m_0m_1}{\mu^2}+\f{m_0^2-m_1^2}{p_1^2}\log \f{m_1}{m_0}-\f{m_0m_1}{p_1^2}\L\f{1}{r}-r\R\log r, 
\end{align} 
where $r$ and $1/r$ are determined by 
\begin{align}
x^2+\f{m_0^2+m_1^2-p_1^2-i\epsilon}{m_0m_1}x+1=(x+r)(x+1/r).
\end{align} 
It has two limits of interest in this paper. Let us consider the first limit, i.e., small external momentum $p_1^2=Q^2\ra0$, then we have
\begin{align}
B_0(Q,m_1,m_2)\doteq  &  \f{x_1+x_2}{2(x_1-x_2)}\log\f{x_2}{x_1}    +\f{1}{2}\log (x_1x_2)-\L\log M^2 +1\R\cr
&+\f{Q^2}{M^2}\f{x_1x_2\L x_1^2-x_2^2+2x_1x_2\log\f{x_2}{x_1} \R}{2(x_1-x_2)^3},
\end{align} 
up to irrelevant additive constants that will be cancelled in the expressions. Here $x_i\equiv M^2/m_i^2$ with $M$ a referred scale. If $x_1=x_2\equiv x$, one can greatly simplify it into 
\begin{align}
B_0(Q,m_1,m_1)\doteq   \log x-\L 2+\log M^2 \R+\f{Q^2}{M^2}\f{x}{6}.
\end{align}
Now we move to the other limit, i.e., { when} one particle { in the loop} is extraordinarily lighter than other mass scales; without loss of generality, let $m_1^2\ll m_2^2,\,p_2^2$. Then one has ($x_2\equiv p_2^2/m_2^2$)
\begin{align}
B_0(p_2,m_1,m_2)\doteq     \f{1}{x_2} \log(1-x_2)  -2{\rm arctanh}(1-2x_2) +\f{1}{2}  \log p_2^2.
\end{align} 

\section{radiative corrections on Higgs-DM-DM vertex}

In this appendix we derive the approximations of Higgs-DM-DM vertex relevant to DM direct detection. The amplitude is given by ${\cal M}+{\cal M}_c$ with
\begin{align}\label{}
{\cal M}\approx -\bar u(p_1-p_2)  A_{ij} \L \ld_{ja}^{L}  P_L+ \ld_{ja}^{R}  P_R \R \gamma^\mu C_{\mu}(-p_2,p_1-p_2)_{aij} 
\L \ld_{ia}^{L*}  P_R+ \ld_{ia}^{R*}  P_L \R  v(p_2),
\end{align} 
where terms suppressed by lepton masses are neglected. For short, we denote $C_{\mu}(-p_2,p_1-p_2,m_{l_a},m_{\wt e_i},{m_{\wt e_j}})\equiv C_{\mu}(-p_2,p_1-p_2)_{aij}$. Similar conventions are adopted throughout this paper. It does not cause confusion since we have specified an unique index type for each flavor. The vectorial three-point function can be decomposed into
\begin{align}\label{}
 \gamma^\mu C_{\mu}(-p_2,p_1-p_2)_{aij} = -\mbox{$\not \hspace{-0.10cm} p_2$ }  C_{11}(-p_2,p_1-p_2)_{aij} +
 (\mbox{$\not \hspace{-0.10cm} p_1$ } -\mbox{$\not \hspace{-0.10cm} p_2$ } )  C_{12}(-p_2,p_1-p_2)_{aij}.
\end{align}
After using the motion of equation, one has $\mbox{$\not \hspace{-0.10cm} p_2$ }\ra -M $ and  $\mbox{$\not \hspace{-0.10cm} p_1$ } -\mbox{$\not \hspace{-0.10cm} p_2$ } \ra +M $. Then, the amplitude takes the form of ${\cal M}\approx -\bar u(p_1-p_2)\L H_L P_L+H_R P_R \R  v(p_2)$ with
\begin{align}\label{}
H_L=\f{1}{16\pi^2}M A_{ij} \left[  \ld_{ia}^{L*}\ld_{ja}^{L}  C_{11}(-p_2,p_1-p_2)_{aij}  +  \ld_{ia}^{R*}\ld_{ja}^{R}  C_{12}(-p_2,p_1-p_2)_{aij}   \right].
\end{align} 
$H_R$ is obtained by exchanging $C_{11}$ and $C_{12}$ in $H_L$. Specific to the kinematics in this paper, i.e., $p_2^2=(p_1-p_2)^2=M^2$, and using the equations below Eq.~(\ref{B0}) and  Eq.~(\ref{C0:1}) one can explicitly show $H_L=H_R$. After some exercise one finds the crossed diagram gives ${\cal M}_c= {\cal M}$. Therefore, eventually the form factor relevant to direct detection is $\ld_h(0)\equiv 2(H_L+H_R)$. In the $\theta\ra0$ limit, the leading order is
\begin{align}\label{}
\ld_{hN}(0)&\approx \sin\theta\f{|y_{La}|^2+|y_{Ra}|^2}{32\pi^2}\f{\mu}{\sqrt{2}M}\left[     B_0(p_1-p_2)_{a1}- B_0(p_1-p_2)_{a2}-2 B_0(p_1)_{11} +2 B_0(p_1)_{12}
 \right.& \nonumber \\
&\left.
-C_0(-p_2,p_1-p_2)_{a12} \L m_{\wt e_1}^2+m_{\wt e_2}^2-2M^2\R+2C_0(-p_2,p_1-p_2)_{a11}\L m_{\wt e_1}^2-M^2\R   \right].
\end{align}  
Note that both the quartic and logarithmic divergencies contained in the two-point functions are cancelled. This is consistent with expectancy and provides as a check for our calculations. It is convenient to write $C_0(-p_2,p_1-p_2)_{a12}={\cal G}(x_1,x_2)/M^2$ with
\begin{align}\label{g12}
{\cal G}(x_1,x_2)\approx -\int_{0}^{1}dx\int_0^x  dy  \f{1}{ y^2-(x_1+1) y+(x_1-x_2)x+x_2},
\end{align}
with  $x_i=m_{\wt e_i}^2/M^2$. It, again, is in the approximation $p_1^2\ra0$ and $m_{\l_a}^2\ra0$;. It has an explicit but not illustrative expression, thus not given here.



\begin{thebibliography}{99}
\itemsep 0.5mm


\bibitem{Bernstein:2013hba} 
  R.~H.~Bernstein and P.~S.~Cooper,
  Phys.\ Rept.\  {\bf 532}, 27 (2013)
  [arXiv:1307.5787 [hep-ex]].
  
  \bibitem{HLFV:early} 
J. Bjorken and S. Weinberg, Phys.Rev.Lett. 38, 622 (1977).

\bibitem{Harnik:2012pb} 
  R.~Harnik, J.~Kopp and J.~Zupan,
  JHEP {\bf 1303}, 026 (2013)
  [arXiv:1209.1397 [hep-ph]].

\bibitem{Khachatryan:2015kon} 
  V.~Khachatryan {\it et al.} [CMS Collaboration],
  Phys.\ Lett.\ B {\bf 749}, 337 (2015). 


\bibitem{LFV:ATLAS} 
  G.~Aad {\it et al.} [ATLAS Collaboration],
  arXiv:1508.03372 [hep-ex].


  \bibitem{Kopp:2014rva} 
  J.~Kopp and M.~Nardecchia,
  JHEP {\bf 1410}, 156 (2014)
  [arXiv:1406.5303 [hep-ph]].

\bibitem{Pilaftsis} 
A. Pilaftsis, Phys. Lett. B 285 (1992) 68.

\bibitem{Arganda:2004bz} 
  E.~Arganda, A.~M.~Curiel, M.~J.~Herrero and D.~Temes,
  Phys.\ Rev.\ D {\bf 71}, 035011 (2005). 
  
  
\bibitem{Arganda:2014dta} 
  E.~Arganda, M.~J.~Herrero, X.~Marcano and C.~Weiland,
  Phys.\ Rev.\ D {\bf 91}, no. 1, 015001 (2015)
  [arXiv:1405.4300 [hep-ph]]. 
  
  \bibitem{Arganda:2015naa} 
  E.~Arganda, M.~J.~Herrero, X.~Marcano and C.~Weiland,
  arXiv:1508.04623 [hep-ph].

  
  
\bibitem{Loop:nu1}
A. Zee, Phys. Lett. B 93, 389 (1980) [Erratum-ibid. B 95, 461 (1980)]; A. Zee, Nucl. Phys. B 264, 99 (1986); K. S. Babu, Phys. Lett. B 203, 132 (1988).
 
\bibitem{Loop:nu2}
E. Ma, Phys. Rev. D 73, 077301 (2006) [arXiv:hep-ph/0601225]. 

\bibitem{Vicente:2014qea} 
D. Aristizabal Sierra and A. Vicente, Phys. Rev. D 90, no. 11, 115004 (2014). 

\bibitem{Heeck:2014qea} 
  J.~Heeck, M.~Holthausen, W.~Rodejohann and Y.~Shimizu,
  Nucl.\ Phys.\ B {\bf 896}, 281 (2015).

\bibitem{Mao:2015hwa} 
  Y.~n.~Mao and S.~h.~Zhu,
  arXiv:1505.07668 [hep-ph].

\bibitem{Omura:2015nja} 
  Y.~Omura, E.~Senaha and K.~Tobe,
  JHEP {\bf 1505}, 028 (2015)
  [arXiv:1502.07824 [hep-ph]].



\bibitem{Dorsner:2015mja} 
  I.~Dorsner, S.~Fajfer, A.~Greljo, J.~F.~Kamenik, N.~Kosnik and I.~Nisandzic,
  arXiv:1502.07784 [hep-ph].

\bibitem{Crivellin:2015mga} 
  A.~Crivellin, G.~D'Ambrosio and J.~Heeck,
  Phys.\ Rev.\ Lett.\  {\bf 114}, no. 15, 151801 (2015). 

\bibitem{Botella:2015hoa} 
  F.~J.~Botella, G.~C.~Branco, M.~Nebot and M.~N.~Rebelo,
  arXiv:1508.05101 [hep-ph].

\bibitem{Campos:2014zaa} 
  M.~D.~Campos, A.~E.~C.~Hern\'andez, H.~P{\"a}s and E.~Schumacher,
  Phys.\ Rev.\ D {\bf 91}, no. 11, 116011 (2015)
  [arXiv:1408.1652 [hep-ph]].

\bibitem{deLima:2015pqa} 
  L.~de Lima, C.~S.~Machado, R.~D.~Matheus and L.~A.~F.~do Prado,
  arXiv:1501.06923 [hep-ph].



\bibitem{Liu:2015oaa} 
  X.~Liu, L.~Bian, X.~Q.~Li and J.~Shu,
  arXiv:1508.05716 [hep-ph].

\bibitem{Crivellin:2015hha} 
  A.~Crivellin, J.~Heeck and P.~Stoffer,
  arXiv:1507.07567 [hep-ph].


\bibitem{Chiang:2015vpt} 
C. W. Chiang, H. Fukuda, M. Takeuchi and T. T. Yanagida, arXiv:1507.04354 [hep-ph].

\bibitem{Huang:2015vpt} 
  W.~Huang and Y.~L.~Tang,
  arXiv:1509.08599 [hep-ph].

\bibitem{He:2015rqa} 
  X.~G.~He, J.~Tandean and Y.~J.~Zheng,
  JHEP {\bf 1509}, 093 (2015)
  [arXiv:1507.02673 [hep-ph]].

\bibitem{Altmannshofer:2015esa} 
  W.~Altmannshofer, S.~Gori, A.~L.~Kagan, L.~Silvestrini and J.~Zupan,
  arXiv:1507.07927 [hep-ph].




\bibitem{Cheung:2015yga} 
  K.~Cheung, W.~Y.~Keung and P.~Y.~Tseng,
  arXiv:1508.01897 [hep-ph].


\bibitem{Baek:2015mea} 
  S.~Baek and K.~Nishiwaki,
  arXiv:1509.07410 [hep-ph].
  
  
  \bibitem{Bi:2009md} 
  X.~J.~Bi, P.~H.~Gu, T.~Li and X.~Zhang,
  JHEP {\bf 0904}, 103 (2009)
  [arXiv:0901.0176 [hep-ph]].


\bibitem{Lee:2014rba} 
  C.~J.~Lee and J.~Tandean,
  JHEP {\bf 1504}, 174 (2015)
  [arXiv:1410.6803 [hep-ph]].


\bibitem{Agrawal:2015tfa} 
  P.~Agrawal, Z.~Chacko, C.~Kilic and C.~B.~Verhaaren,
  arXiv:1503.03057 [hep-ph].
  
 \bibitem{Hamze:2014wca} 
  A.~Hamze, C.~Kilic, J.~Koeller, C.~Trendafilova and J.~H.~Yu,
  Phys.\ Rev.\ D {\bf 91}, no. 3, 035009 (2015)
  [arXiv:1410.3030 [hep-ph]].

\bibitem{Bai:2014osa} 
  Y.~Bai and J.~Berger,
  JHEP {\bf 1408}, 153 (2014)
  [arXiv:1402.6696 [hep-ph]].
  
  \bibitem{Geng:2014zqa} 
  C.~Q.~Geng, D.~Huang and L.~H.~Tsai,
  JHEP {\bf 1408}, 086 (2014)
  [arXiv:1406.6481 [hep-ph]].


\bibitem{Kile:2014jea} 
  J.~Kile, A.~Kobach and A.~Soni,
  Phys.\ Lett.\ B {\bf 744}, 330 (2015)
  [arXiv:1411.1407 [hep-ph]].
  
 \bibitem{Agrawal} 
P. Agrawal, S. Blanchet, Z. Chacko, and C. Kilic, Phys. Rev. D86 (2012) 055002. 


\bibitem{Chang:2014tea} 
  S.~Chang, R.~Edezhath, J.~Hutchinson and M.~Luty,
  Phys.\ Rev.\ D {\bf 90}, no. 1, 015011 (2014)
  [arXiv:1402.7358 [hep-ph]].


\bibitem{Akerib:2013tjd} 
  D.~S.~Akerib {\it et al.}  [LUX Collaboration],
  Phys.\ Rev.\ Lett.\  {\bf 112}, 091303 (2014). 


\bibitem{Denner:1991kt} 
  A.~Denner,
  Fortsch.\ Phys.\  {\bf 41}, 307 (1993)
  [arXiv:0709.1075 [hep-ph]].

\bibitem{tau:mugamma} 
B. Aubert et al. (BaBar Collaboration), Phys.Rev.Lett., 104, 021802 (2010). 

\bibitem{tau:mugamma1} 
T. Aushev, W. Bartel, A. Bondar, J. Brodzicka, T. Browder, et al., (2010), arXiv:1002.5012 [hep-ex].


\bibitem{CLFV:mu} 
Adam et al. (MEG Collaboration), Phys.Rev.Lett. 110, 201801 (2013), 1303.0754.


\bibitem{hFF} 
D. Carmi, A. Falkowski, E. Kuflik, T. Volansky and J. Zupan, JHEP 1210, 196 (2012); 
  J.~Guo, Z.~Kang, J.~Li and T.~Li,
  arXiv:1308.3075 [hep-ph].


\bibitem{Bernon:2014vta} 
  J.~Bernon, B.~Dumont and S.~Kraml,
  Phys.\ Rev.\ D {\bf 90}, 071301 (2014). 


\bibitem{Garny:2015wea} 
  M.~Garny, A.~Ibarra and S.~Vogl,
  arXiv:1503.01500 [hep-ph].


 \bibitem{coann}  
  K. Griest and D. Seckel, Phys. Rev. D43 (1991) 3191.
  
  

\bibitem{X. Gao} 
X. Gao, Z. Kang and T. Li, JCAP 1301, 021 (2013).


\bibitem{Alarcon:2011zs} 
  J.~M.~Alarcon, J.~Martin Camalich and J.~A.~Oller,
  Phys.\ Rev.\ D {\bf 85}, 051503 (2012); 
  A.~Crivellin, M.~Hoferichter and M.~Procura,
  Phys.\ Rev.\ D {\bf 89}, 054021 (2014). 



 \bibitem{anpole} 
 B. Kayser and A. S. Goldhaber, Phys.Rev. D28, 2341
(1983).


\bibitem{Kopp} 
  J.~Kopp, L.~Michaels and J.~Smirnov,
  JCAP {\bf 1404}, 022 (2014). 
  
\bibitem{oneloop} 
G. ?t Hooft and M. Veltman, 
Nucl.Phys. B153 (1979) 365-401.









\end{thebibliography}
\end{document}